\documentstyle[aps,preprint,prb]{revtex}

\title{Tunneling Between Parallel Two-Dimensional Electron Gases}

\author{N. Turner \and J. T. Nicholls \and E. H. Linfield 
\and K. M. Brown \and G. A. C. Jones \and D. A. Ritchie}

\address{Cavendish Laboratory, Madingley Road, Cambridge CB3 0HE, UK}

\date{\today}

\begin{document}

\tighten

\widetext

\maketitle

\begin{abstract}

The tunneling between two parallel two-dimensional electron gases has
been investigated as a function of temperature $T$, carrier density
$n$, and the applied perpendicular magnetic field $B$.  In zero
magnetic field the equilibrium resonant lineshape is Lorentzian,
reflecting the Lorentzian form of the spectral functions within each
layer.  From the width of the tunneling resonance the lifetime of
the electrons within a 2DEG has been measured as a function of $n$ and
$T$, giving information about the density dependence of the
electron-impurity scattering and the temperature dependence 
of the electron-electron scattering.
In a magnetic field there is a general suppression of equilibrium
tunneling for fields above $B=0.6$~T.  A gap in the tunneling
density of states has been measured over a wide range of magnetic
fields and filling factors, and various theoretical predictions have
been examined.  In a strong magnetic field, when there is only one
partially filled Landau level in each layer, the temperature
dependence of the conductance characteristics has been modeled with a
double-Gaussian spectral density.

\end{abstract}

\pacs{73.20.Dx 73.40.Gk 73.50.Bk 73.50.Jt} 

\begin{section}{Introduction}
\label{s:intro}

In addition to conserving energy $E$, resonant tunneling between two
parallel two-dimensional electron gases (2DEGs) requires\cite{smol89}
the conservation of the in-plane momentum $\hbar \bbox{k}$.  In zero
magnetic field, the conservation of both quantities allows the
investigation of the broadening of the electronic states within an
individual 2DEG, in a way that is not possible with conventional
transport measurements.  
For example, recent experimental studies\cite{murp95} have shown
that the lifetime of the two-dimensional electrons 
can be measured directly by tunneling spectroscopy. 
The lifetime broadening is characterized by
the electron spectral function $A(\bbox{k},E)$, which describes the
probability that an electron in a particular $\bbox{k}$-state has
energy $E$.  In general, $A(\bbox{k},E)$ is strongly peaked near the
free-particle energy $E_{\bbox{k}}=\hbar^2 k^2 / 2m^\star$, with a
width $\Gamma$ determined by the lifetime $\tau$ of the electron.

In a strong perpendicular magnetic field, when the Landau level
filling factor $\nu = hn/eB$ is less than unity, it is assumed that
the electron traverses the tunnel barrier between the two 2DEGs in a
much shorter timescale than the charge rearrangement around the
injected electron (and around the hole left behind in the emitter
layer).  The energies of the injection and extraction processes create
a gap in the tunneling density of states (DOS).  Many-body theories
predict\cite{yang93,hats93,song93a} that for a fixed filling factor,
$\nu < 1$, the width of the energy gap is approximately $0.4e^2/4\pi
\epsilon l_B$ (where $l_B=\sqrt{\hbar/eB}$ is the magnetic length), or
if the electron liquid forms a regular lattice\cite{johan93} the gap
energy is $e^2/4\pi \epsilon a$ (where $a=1/\sqrt{n\pi}$ is the
inter-electron spacing within a layer).  At fixed filling factor both
predicted forms for the gap have the same $\sqrt{B}$ dependence with
magnetic field.  In an earlier paper we reported\cite{brown94c}
tunneling characteristics in the high magnetic field regime ($\nu<1$).
The theoretical predictions do not match the measured linear field
dependence, $\Delta \approx 0.44 \hbar \omega_{\text{c}}$ 
(where $\hbar\omega_{\text{c}}=\hbar eB/m^\star$ is the cyclotron energy),
that we have observed\cite{brown94c,brown94e} in high magnetic fields
both at fixed filling factor and constant carrier density.

From tunneling measurements of 2D-2D samples with different barrier
widths, Eisenstein {\em et al.\/}\cite{eis95} argue that the high
field gap energy is largely determined by the in-plane Coulomb energy,
modified by an interlayer excitonic binding energy.  The form of the
gap used to fit their data was $\Delta_{\text{E}} = f(\nu)\,
e^2/4\pi\epsilon a - 0.4\, e^2/4\pi\epsilon d$, where $f(\nu)$ is some
universal function of filling factor, and $d$ is the quantum well
center-to-center separation.  Although the interpretation of our data
is different to that of Eisenstein {\em et al.\/}\cite{eis95} (there
is no universal $f(\nu)$ which allows us to plot our high field data
in this form), we measure gaps of comparable energy in lower mobility,
but otherwise similar GaAs/AlGaAs double-layer structures.

In a 2D-3D tunneling system Ashoori {\em et al.\/}\cite{ash93}
measured a filling factor independent suppression of tunneling for
$\nu > 1$.  The temperature dependence of the conductance was fitted
to a model with a linear variation of the DOS about the Fermi energy
$E_{\text{F}}$.  The half-width at half-depth of the gap was measured
to be linear in magnetic field, with a dependence $\Delta_{\text{A}}
\approx \hbar \omega_{\text{c}}/20$.  Taking into account the
difference in the two tunneling systems and the definitions of
$\Delta_{\text{A}}$ and $\Delta$, the two gap widths can be
argued\cite{ashpriv} to be comparable.

Here we consolidate earlier work\cite{brown94c,brown94e,brown94d} at
zero and high magnetic fields, and present data taken at intermediate
fields.  The rest of this paper is organized as follows.  The samples
and measurement details are described in Sec.~\ref{s:exp}.  The basic
tunneling formalism is reviewed in Sec.~\ref{s:formal}, showing how
the tunneling conductance is determined by the underlying electron
spectral functions within each 2DEG.  In Sec.~\ref{s:zerob} we present
tunneling measurements in zero magnetic field, investigating the
various scattering mechanisms which are responsible for the linewidth
of the resonance.  The tunneling properties in a perpendicular
magnetic field are investigated in Secs.~\ref{s:perpb}
and~\ref{s:temp}.

\end{section}

\begin{section}{Experimental Details}
\label{s:exp}

The measurements presented in this paper were obtained from two
samples, C751 and C770, each consisting of two modulation-doped
180~\AA\ wide GaAs quantum wells separated by a 100~\AA\ (C770) or
125~\AA\ (C751) $\text{Al}_{0.33}\text{Ga}_{0.67}\text{As}$ barrier.
The double-well structures were grown by molecular beam epitaxy on
an $\text{n}^+$ epilayer, which had previously been patterned by a
focussed Ga ion beam to form backgates.\cite{brown94a} Optical
lithography was used to fabricate a mesa, Ohmic contacts, and to
define surface Schottky gates.  Small front (back) gates near the
Ohmic contacts were used to selectively deplete the upper (lower)
2DEG, forming an independent contact to the lower (upper) layer.  The
two 2DEGs overlap in a $100~\mu$m$ \times 150~\mu$m tunneling region.
The as-grown carrier densities of the upper and lower layers in the
tunneling region are $n_1 \approx 3 \times 10^{11}$~cm$^{-2}$ and $n_2
\approx 2 \times 10^{11}$~cm$^{-2}$.  The corresponding, as-grown,
low-temperature, mobilities of the two layers are $\mu_1 \approx 8
\times 10^5$~cm$^2$/Vs and $\mu_2 \approx 2 \times
10^5$~cm$^2$/Vs.  The carrier densities in the tunneling region
are independently controlled by voltages $V_{\text{g}_1}$ and
$V_{\text{g}_2}$ applied to large area front- and backgates.

Two-terminal conductance measurements
of the sequential depopulation of Landau levels (LLs)
in each layer were used to determine the relationship between
$V_{\text{g}_{1,2}}$ and the corresponding carrier density $n_{1,2}$.
For example, in C770 the top layer has a density
$n_1 = (7.1 V_{\text{g}_1}/\text{V} + 2.9) 
\times 10^{11}$~cm$^{-2}$,
where $-0.4\leq V_{\text{g}_1}\leq 0.35$~V.
The carrier density at $V_{\text{g}_1}=0$ can vary by as much
as 10\% from one sample cooldown to another.
By contrast, the constant of proportionality (7.1)
between $V_{\text{g}_1}$ and $n_1$ is always the same,
and is in good agreement with a model where the 
top 2DEG and the front gate form a parallel 
plate capacitor of separation 1000~\AA.

The differential tunneling conductance $G=dI/dV_{\text{sd}}$ was
measured using constant voltage (0.1~mV) ac lockin techniques
as a function of carrier density $n$, temperature $T$, and applied dc
interlayer source-drain voltage $V_{\text{sd}}$.  For measurements in
a perpendicular magnetic field $B$ the carrier densities were always
matched, and the given values of $\nu$ refer to the filling factors in
each layer.

\end{section}

\begin{section}{Tunneling Formalism}
\label{s:formal}

Tunneling accesses the single-particle DOS (the DOS available for the
addition of another electron), compared to conventional transport
measurements which provide information about the DOS available for the
excitation of an electron from the Fermi sea into a conducting state.
Zheng and MacDonald\cite{zheng93a} showed theoretically that 2D-2D
tunneling characteristics in zero magnetic field can be used to
determine the spectral function of an electron.  Within layer $i$
[$i=1,2$ (upper, lower)] scattering introduces an energy broadening
$\Gamma_i = \hbar/2 \tau_i$, where $\tau_i$ is the scattering time.
In the Born approximation the spectral function within each 2DEG will
have the Lorentzian form\cite{mahan93}
\begin{equation}\label{e:lorentz}
A_i(\bbox{k},E) \propto \frac{1}{\Gamma_i^2 +
\left(\frac{\hbar^2 k^2}{2m^\star} - E \right)^2},
\end{equation}
where $E$ is measured from the bottom of the 2D subband.
Using standard notation\cite{wolf85}
the tunneling current $I(V_{\text{sd}})$ 
between the two layers is calculated
from the two spectral functions to be
\begin{eqnarray}\label{e:current}
I(V_{\text{sd}}) & \propto & \int d^2\bbox{k}_1 \int d^2\bbox{k}_2 \;
\mid\! T_{\bbox{k}_1, \: \bbox{k}_2} \!\mid^2 \int dE \nonumber\\
&& A_1(\bbox{k}_1,E-E_{0_1})\, A_2(\bbox{k}_2,E-E_{0_2}) \nonumber\\
&& \times [f(E-\mu_1,T) - f(E-\mu_2,T)],
\end{eqnarray}
where  $f(E,T)=1/(1+\exp(E/k_{\text{B}}T))$ 
is the Fermi distribution function.
The chemical potentials $\mu_i$ and
the 2D subband energies $E_{0_i}$ 
are defined in Fig.~\ref{f:band}.

The integral is simplified by the following assumptions:
the absolute conservation of momentum 
(such that the interlayer tunneling matrix element
$T_{\bbox{k}_1, \: \bbox{k}_2}$ is proportional to
$\delta_{\bbox{k}_1, \: \bbox{k}_2}$),
low temperature ($k_{\text{B}}T \ll \Gamma$, 
with Fermi functions approximated by step functions),
and weak disorder ($\Gamma \ll E_{\text F}$).
Using these assumptions,
substitution of Lorentzian spectral functions
into Eq.~\ref{e:current} gives the tunneling current
\begin{equation}\label{e:current2}
I(V_{\text{sd}}) = \frac{G_0 \Gamma^2}
{\Gamma^2+(\Delta\! E_{\text{F}}+ eV_{\text{sd}})^2}  V_{\text{sd}},
\end{equation}
where $\Delta\! E_{\text{F}}=E_{\text{F}_2}-E_{\text{F}_1}$,
and $G_0$ is a constant with the dimensions of conductance.
The tunneling linewidth is the sum, $\Gamma =\Gamma_1+\Gamma_2$,
of the widths of the Lorentzian 
spectral functions in the two layers.\cite{zheng93a}

In this paper we report differential
conductance $G= dI/dV_{\text{sd}}$ measurements.
The zero-bias conductance obtained 
by differentiating Eq.~\ref{e:current} is
\begin{equation}\label{e:conductance}
G(V_{\text{sd}}=0) \propto \int d^2\bbox{k} \; 
A_1(\bbox{k},E_{\text{F}_1}) \, A_2(\bbox{k},E_{\text{F}_2}).
\end{equation}
Using Lorentzian spectral functions,
the differential conductance as one subband edge 
(controlled by a gate voltage $V_{\text{g}}$)
is swept past the other is\cite{zheng93a}
\begin{equation}\label{e:conductance2}
G(V_{\text{g}}) = G_0 \frac{\Gamma^2}{\Gamma^2 + \Delta\! E_{\text{F}}^2}.
\end{equation}
Equation~\ref{e:conductance2} predicts  
Lorentzian $G(V_{\text{g}})$ characteristics
with a linewidth $\Gamma$ and peaked with
magnitude $G_0$ at $\Delta\! E_{\text{F}}=0$,
when the carrier densities in the two wells are equal.

We are interested in the differential 
conductance as a function of $V_{\text{sd}}$,
taken when the carrier densities have been matched at $V_{\text{sd}}=0$. 
However, upon application of an interlayer dc voltage
the capacitance between the two layers unavoidably
mismatches the carrier densities,
and it is {\em incorrect\/} to set
$\Delta\! E_{\text{F}} = 0$ in the derivative of 
Eq.~\ref{e:current2} to obtain the conductance 
\begin{equation}\label{e:noneqbm1}
G(V_{\text{sd}}) =G_0 \frac{\Gamma^2(\Gamma^2-(eV_{\text{sd}})^2)}
{\left(\Gamma^2+(eV_{\text{sd}})^2\right)^2}.
\end{equation} 
There are in fact two contributions to $\Delta\!
E_{\text{F}}$:  One component $\Delta\! E_{\text{F}}^0$ is due to the
applied gate voltages (this is zero when the carrier densities are
matched at $V_{\text{sd}}=0$), and the other, $\Delta\!
E_{\text{F}}^{\text{C}}$, is a capacitive effect.  $\Delta\!
E_{\text{F}}^{\text{C}}$ is determined by the change of carrier
density $\delta\! n = \pm C V_{\text{sd}}/e$ induced by
$V_{\text{sd}}$, where $C = \epsilon/d$ is the capacitance per unit
area between the two 2DEGs.  The capacitive change in $\Delta\!
E_{\text{F}}$ is $\Delta\! E_{\text{F}}^{\text{C}} = -2\mid\!\delta\!
n\!\mid / \rho_0 = -\beta eV_{\text{sd}}$, where $\rho_0 =
m^{\star}/\pi \hbar^2$ is the 2D density of states, and $\beta =
a_{\text{B}}/2d$ (where the Bohr radius is
$a_{\text{B}}\approx100$~\AA\ in GaAs).  Substitution of
$\Delta\! E_{\text{F}} = \Delta\! E_{\text{F}}^0 + \Delta\!
E_{\text{F}}^{\text{C}} = \Delta\!E_{\text{F}}^0 - \beta
eV_{\text{sd}}$ into Eq.~\ref{e:current2}, followed by
differentiation, gives the modified lineshape
\begin{equation}
G(V_{\text{sd}}) = G_0 \frac{\Gamma^2(\Gamma^2+(\Delta\!
E_{\text{F}}^0)^2-(eV_{\text{sd}}(1-\beta))^2)}
{\left(\Gamma^2+(\Delta\! E_{\text{F}}^0 + eV_{\text{sd}}(1-\beta))^2
\right)^2}.
\label{e:noneqbmgen}
\end{equation}
Setting $\Delta\! E_{\text{F}}^0 =0$ 
(for matched carrier densities at $V_{\text{sd}}=0$)
gives the lineshape 
\begin{equation}
G(V_{\text{sd}}) =G_0 
\frac{\Gamma^2(\Gamma^2-(eV_{\text{sd}}(1-\beta))^2)} 
{\left(\Gamma^2+(eV_{\text{sd}}(1-\beta))^2\right)^2},
\label{e:noneqbm2}
\end{equation}
which has a linewidth of $\Gamma/(1-\beta)$,
in contrast to the linewidth $\Gamma$ 
given in Eq.~\ref{e:noneqbm1}.
A typical value of $\beta$ is 0.15.
Therefore, linewidths obtained from $G(V_{\text{sd}})$
measurements are approximately 18\% larger than those
obtained from equilibrium conductance measurements.\cite{brown94d} 
Similar capacitive corrections should also be applied to linewidths
obtained\cite{murp95} from plots of $I/V_{\text{sd}}$ versus $V_{\text{sd}}$.

We can only quantify capacitive effects at $B=0$,
though the magnitude of such effects in a 
perpendicular magnetic field can be estimated 
by replacing $\rho_0$ with $\rho_B$,
the density of states at the Fermi level in the outer Landau level.
If the outer LL is half-filled,
$\rho_B \gg \rho_0$,
then the capacitive change of the Fermi energies 
$\Delta\! E_{\text{F}}^{\text{C}}$ will be negligible.
The opposite result is obtained near integral filling factors
and so we have not made any conclusions from 
tunneling measurements in this regime.

In Sec.~\ref{s:temp} we model high field tunneling 
measurements obtained at filling factor $\nu=1/2$.
To model the data we have guessed the form of the spectral density 
$A_i(E) = 1/(2\pi)^2 \int d^2\bbox{k}\; A_i(\bbox{k},E)$,
and have calculated the tunneling current using the expression 
\begin{eqnarray}
I(V_{\text{sd}}) & \propto & \int dE\; A_1(E-E_{0_1})\, A_2(E-E_{0_2})
\nonumber\\
&& \times [f(E-\mu_1,T) - f(E-\mu_2,T)].
\label{e:spect}
\end{eqnarray}

\end{section}

\begin{section}{Tunneling in Zero Magnetic Field}
\label{s:zerob}

Figure~\ref{f:lorentz} shows the zero-field differential tunneling
conductance $G$ as a function of $V_{\text{g}_1}$, the front gate voltage;
by fixing $V_{\text{g}_2}$, the carrier density in the bottom 
layer was held constant at $n_2=3.25\times 10^{11}$~cm$^{-2}$. 
The equilibrium measurements were taken
at $T=3.0$~K (squares) and 19.0~K (circles).
Due to a misalignment between the 
frontgate and backgate in the tunneling region,
there is an unwanted second resonance 
(which is independent of $V_{\text{g}_2}$),
as well as a small constant background conductance. 
In Fig.~\ref{f:lorentz} both of these weak contributions have been removed,
and the corrected data at the two temperatures 
have been fitted to the Lorentzian lineshape
\begin{equation}
G(V_{\text{g}_1})=\frac{G_0}{1+\left(\frac{V_{\text{g}_1}-V_0}
{\delta\! V_{\text{g}}}\right)^2},
\label{e:lorentz2}
\end{equation}
which is Eq.~\ref{e:conductance2} rewritten explicitly in terms of the
front gate voltage $V_{\text{g1}}$.  The fits, shown as solid lines, are with
$V_0=0.046$~V, $\delta\! V_{\text{g}}=0.0194$~V, and
$G_0=16.1~\mu$S at 3~K, and with $V_0=0.045$~V,
$\delta\! V_{\text{g}}=0.0360$~V, and $G_0=9.9~\mu$S
at 19~K.  We have also tried fitting the resonances to a
Gaussian and to the derivative of a Fermi function ($f^\prime \sim
\text{sech}^2 [(V_{\text{g}_1}-V_0)/\delta\! V_{\text{g}}]$).  Of the
three functional forms, a Gaussian fit is the poorest, $f^\prime$ is
considerably better, but the overall shape and the tails of the
resonance are best fit by a Lorentzian.  The half width at half
maximum of the low-temperature Lorentzian resonance,
$\delta\!V_{\text{g}} \approx 19$~mV, corresponds to an energy
broadening $\Gamma \approx 0.5$~meV.  This linewidth $\Gamma$ is the
sum of the widths of the spectral functions in the double-layer
system, $\Gamma = \hbar/2\tau_1 + \hbar/2\tau_2$.  We relate the
linewidth to the scattering rate, defining the average scattering rate
of the double-layer system to be
$\tau^{-1}=(\tau^{-1}_1+\tau^{-1}_2)/2=\Gamma/\hbar$.

Non-equilibrium measurements provide an alternative method for
determining the linewidth $\Gamma$.  Figure~\ref{f:betafig} shows a
comparison of the non-equilibrium $G(V_{\text{sd}})$ (squares) and
equilibrium $G(V_{\text{g}_1})$ characteristics (circles, and offset
vertically) for the same $n_2$ for sample C751.  To demonstrate the
validity of Eq.~\ref{e:noneqbmgen}, the non-equilibrium data was
obtained with a small mismatch of carrier densities, causing a 
slight shift of the central $G(V_{\text{sd}})$ peak to the
left of $V_{\text{sd}}=0$.  The solid line through the
$G(V_{\text{sd}})$ data is a fit to Eq.~\ref{e:noneqbmgen} with
$\Delta\! E_{\text{F}}^0 = 0.08$~meV, corresponding to a carrier
density mismatch of 0.7\%.  Though small, this mismatch is enough to
cause a noticeable asymmetry in the $G(V_{\text{sd}})$
characteristics.  The fits to both the equilibrium and non-equilibrium
traces in Fig.~\ref{f:betafig} give $\Gamma=0.384$~meV and
$G_0=3.03~\mu$S.  
The fitted value $\beta=0.14 \pm 0.01$ 
is obtained from the non-equilibrium trace and corresponds
to an interlayer separation of $d=365\pm 30$~\AA, somewhat
larger than the quantum well center-to-center separation of
305~\AA\ deduced from the growth parameters.  However, both
parallel magnetic field measurements (using the method of
Ref.~\onlinecite{eis91b}) and self-consistent calculations suggest a
center-to-center wavefunction separation of $d\approx340$-350~\AA.

The successful fits to both the
equilibrium and non-equilibrium resonances in Fig.~\ref{f:betafig}
provide strong evidence for a Lorentzian 
spectral function in zero magnetic field.
We have been able to fit Lorentzian lineshapes 
to the equilibrium tunneling resonances
obtained over a wide range of carrier densities,
$0.3\leq n \leq 3.2\times10^{11}$~cm$^{-2}$,
at temperatures from 3~K to 19~K.
From the $n$ and $T$ dependence of the linewidth $\Gamma$,
we have investigated the electron-impurity, 
electron-phonon, and electron-electron contributions
to the overall scattering rate 
\begin{equation}\label{e:rate}
\Gamma/\hbar=\tau^{-1}=\tau^{-1}_{\text{e-imp}}+\tau^{-1}_{\text{e-ph}}+
\tau^{-1}_{\text{e-e}}.
\end{equation}

Figure~\ref{f:widtht} shows the temperature dependence of the
equilibrium linewidth $\Gamma$ at matched carrier densities of
$n=0.91,\, 1.62,\, 2.19,\, 3.04\times 10^{11}$~cm$^{-2}$.  The
solid lines are fits of the form $\Gamma=a+bT^2$; at high $n$ the fits
are good, but at lower densities the measured linewidth departs from
quadratic behavior.  The fitting parameters are both functions of $n$,
with typical values $a \approx 0.5$-0.8~meV and $b \sim 5\times
10^{-3}$~meV/K$^{2}$.  Only the electron-phonon and electron-electron
scattering terms in Eq.~\ref{e:rate} are expected to show significant
temperature dependence.  In the temperature range considered here, the
sheet resistivity $\rho_{xx}$ of each 2DEG shows only a weak
temperature dependence, indicating that the mobility is dominated by
impurity scattering, and that, by comparison, the temperature dependent
electron-phonon contribution is small.  We therefore associate the
temperature dependent component of the tunneling linewidth with that
due to electron-electron scattering $\Gamma_{\text{e-e}}$.  The
electron-electron scattering rate in a clean 2D metal is
given by the well-known theoretical expression\cite{giuliani82} 
\begin{equation}\label{e:gnq}
\frac{\hbar}{\tau_{\text{e-e}}} = \frac{E_{\text{F}}}{2\pi}
\left(\frac{k_{\text{B}}T}{E_{\text{F}}}\right)^2
\left[\ln\left(\frac{E_{\text{F}}}{k_{\text{B}}T}\right)+
\ln\left(\frac{2q_{\text{TF}}}{k_{\text{F}}}\right)+1\right],
\end{equation}
where $q_{\text{TF}}=2/a_{\text{B}}$ is 
the 2D Thomas-Fermi screening wavevector.
At fixed carrier density $\hbar/\tau_{\text{e-e}}$ 
exhibits an approximate $T^2$ temperature dependence,
in agreement with the solid line fits used in Fig.~\ref{f:widtht}. 

At low temperatures, the electron-electron scattering is negligible,
and $\tau^{-1}$ is dominated by electron-impurity scattering.
Figure~\ref{f:logglogn} shows a log-log plot 
of the equilibrium linewidth
versus carrier density at $T=3$~K.
The fitted gradient is $x=-0.47\pm0.05$,
which is approximately the power-law behavior
expected\cite{gold88} for small-angle scattering
from remote ionized impurities,
\begin{equation}\label{e:gold}
\frac{\hbar}{\tau_{\text{e-imp}}}=
\frac{\hbar^2n_{\text{D}}}{2m^\star s}\sqrt{\frac{\pi}{2}}\,n^{-1/2},
\end{equation}
where $n_{\text{D}}$ is the sheet density of impurities,
and $s$ is the spacer layer thickness.
Equation~\ref{e:gold} is valid for $4k_{\text{F}}s \gg 1$, 
where $k_{\text{F}}$ is the Fermi wavevector.
In our samples the ionized donors are distributed over distances of 
400-700~\AA\ from the center of each quantum well.
Taking the smallest distance, $s = 400$~\AA, 
the condition $4k_Fs \gg 1$ is easily
satisfied for $n=3\times 10^{11}$~cm$^{-2}$,
though the applicability of Eq.~\ref{e:gold} becomes 
questionable at the lowest carrier densities.

From experiments over a range of carrier densities and temperatures
we have fitted over 300 measured equilibrium 
linewidths $\Gamma(n,T)$ to the expression
\begin{equation}\label{e:ratent}
\Gamma(n,T)= \Gamma_{\text{e-imp}} + \Gamma_{\text{e-e}}
= a_1\,n^{-0.5} + a_2\times[\text{Eq.\ \ref{e:gnq}}].
\end{equation}
If $\Gamma$ is measured in units of meV, 
$n$ in units of $10^{11}$~cm$^{-2}$, and $T$ in K, 
the best fit was obtained with the numerical values
$a_1=0.77\pm0.01$ and $a_2=3.06\pm0.09$. 
The dashed lines in Fig.~\ref{f:widtht} 
are obtained from this best fit.

The magnitude of the electron-impurity scattering rate,
$\Gamma_{\text{e-imp}} = a_1 n^{-0.5}$ in Eq.~\ref{e:ratent}, is in
agreement with other experiments,\cite{coleridge91,das93} but is
fifteen times smaller than the theoretical prediction of Eq.~\ref{e:gold}.
Discrepancies between theory and experiment have 
previously been resolved\cite{vanhall89} 
by including correlations between scatterers.  In the high $n$, low $T$
regime, the ratio $\tau^{-1}/\tau_{\text{t}}^{-1}\approx 7$ (where
$\tau_{\text{t}}$ is the mobility lifetime) agrees with
measurements\cite{coleridge91} of similar mobility single-layer
GaAs/AlGaAs heterostructures.  Furthermore, the scattering time
$\tau\approx 1.2$~ps is comparable to the Dingle time
1.1~ps (obtained from Shubnikov-de Haas oscillations measured
on a Hall bar made from the same wafer), confirming the importance of
small-angle scattering in determining the tunneling linewidth.

To look at just the electron-electron scattering term,
Fig.~\ref{f:univ} shows 
$\Gamma_{\text{e-e}}/E_{\text{F}}$ plotted versus
$k_{\text{B}}T/E_{\text{F}}$.
This representation of the data is 
expected to be nearly universal,\cite{murp95}
though there is some dependence on carrier density inside the square bracket
of Eq.~\ref{e:gnq}.
This weak dependence on $n$ is illustrated by the
two solid line traces of 3.06$\times$[Eq.~\ref{e:gnq}] (our best
fit for $\Gamma_{\text{e-e}}$) for  $n=0.3$ and
$3\times10^{11}$~cm$^{-2}$ (upper and lower solid curves, respectively).
Equation~\ref{e:gnq} is plotted 
for $n=1.3\times10^{11}$~cm$^{-2}$ and 
without any prefactor ($a_2=1$)
as the dashed line labelled GQ in Fig.~\ref{f:univ}.
Since the original derivation by Giuliani
and Quinn\cite{giuliani82} 
there has been some confusion about the value of $a_2$
that multiplies this expression.  
Fukuyama and Abrahams\cite{fuk83} calculated the expression
\begin{equation}\label{e:fuk}
\frac{\hbar}{\tau_{\text{e-e}}} = \pi^2 \, \frac{E_{\text{F}}}{2\pi}
\left(\frac{k_{\text{B}}T}{E_{\text{F}}}\right)^2
\ln\left(\frac{E_{\text{F}}}{k_{\text{B}}T}\right),
\end{equation}
and this is shown in Fig.~\ref{f:univ}
as the curve labelled FA.\cite{note2}
More recent theories\cite{jung96a,zheng96} 
highlight the mathematical errors in
Ref.~\onlinecite{giuliani82} and suggest that 
due to the nature of the tunneling experiment,
there are both electron and hole contributions to the linewidth.
These two considerations result in $a_2=\pi^2/2=4.9$,
giving the third dashed line in Fig.~\ref{f:univ} labelled JM.
Without any energy dependent corrections 
the various predictions (GQ, FA, and JM) can be compared 
to our equilibrium measurement $a_2 =3.06$.
The difference between the predictions of GQ and JM and our 
fit is purely in the value of the prefactor $a_2$;
whereas the curve FA has a functional form which does not 
describe the data very well. 

The greatest discrepancies between theory
and experiment in Fig.~\ref{f:univ} occur for 
large $k_{\text{B}}T/E_{\text{F}}$,
for which there are various possible reasons.
First, it is assumed in the derivation\cite{giuliani82} of Eq.~\ref{e:gnq}
that $E_{\text{F}} \gg k_{\text{B}}T$,
a condition which breaks down at low densities and high temperatures.
Second, when the matched carrier densities are reduced there
is an increase in the disorder of the two layers,
and the 2DEGs may no longer be considered as clean 2D metals.
The increased level of disorder may also render the Born approximation
invalid, in which case the spectral functions 
and the tunneling resonance are no longer Lorentzian.
Third, the derivation of Lorentzian $G(V_{\text{g}_1})$ 
characteristics (Eq.~\ref{e:lorentz2}) from Lorentzian
spectral functions is only valid when $\Gamma \ll E_{\text{F}}$;
the two energies become comparable when
$n = 0.5\times 10^{11}$~cm$^{-2}$.
The fourth reason is experimental:
the pinch-off characteristics 
of the gate ($ V_{\text{g}_1})$ 
distort the symmetric equilibrium lineshape,
increasing the uncertainty of measured linewidths at low $n$.

In our analysis of the temperature 
dependence of the equilibrium linewidth we have followed an
approach similar to that of Murphy {\em et al.\/},\cite{murp95} 
where the non-equilibrium linewidths were measured from
$I/V_{\text{sd}}$ versus $V_{\text{sd}}$ 
characteristics of double 2DEG structures of higher mobility 
($\mu\approx10^6$~cm$^2$/Vs).
In our samples the measured electron-electron scattering rate
is enhanced over Eq.~\ref{e:gnq} by a factor $a_2 \approx 3$,
which is two times smaller than that 
measured in Ref.~\onlinecite{murp95}.
The two measurements are different, 
and non-equilibrium linewidths are expected to be larger
than equilibrium linewidths for two reasons:
an 18\% enhancement can be attributed 
to charge transfer effects (see Sec.~\ref{s:formal}),
and a further 15\% (Ref.~\onlinecite{jung96a}) 
to 40\% (Ref.~\onlinecite{zheng96}) enhancement
may be introduced by the energy dependence of $\Gamma_{\text{e-e}}$.
In order to achieve an enhancement factor $a_2\approx6$ close to that of 
Murphy {\em et al.\/}\cite{murp95} 
recent theories\cite{jung96a,zheng96} 
have concentrated on the energy dependence
of $\Gamma_{\text{e-e}}$, but have neglected the interlayer
charge transfer effects.
In Fig.~\ref{f:betafig} we show that, in our samples,
we can obtain good agreement between equilibrium 
and non-equilibrium linewidths
by including only charge transfer effects. 
Moreover, in further investigations over a range of 
matched and mismatched carrier densities,
we find that the energy dependent corrections are consistently small.

There are two major differences between our samples and those 
of Ref.~\onlinecite{murp95} 
that might account for the different measured values of $a_2$.
The first is the level of disorder;
in our samples the electron-impurity contribution to the linewidth
is $\Gamma_{\text{e-imp}}\approx 0.5$~meV compared 
to 0.25~meV in Ref.~\onlinecite{murp95}.
This greater disorder probably does not affect the 
applicability of Eq.~\ref{e:gnq},
but will make the energy dependent 
corrections to $\Gamma_{\text{e-e}}$ less important.\cite{zhengpriv}
The second difference is the thickness of the tunnel barrier,
which will affect the strength of the 
interlayer electron-electron interactions.
For samples with quantum well widths of 200~\AA\
and barrier thicknesses in the range 175-340~\AA\ (giving $d=375$-540~\AA),
Murphy {\em et al.\/}\cite{murp95} measure an enhancement $a_2=6.3$
which appears to be independent of barrier thickness. 
In our samples the interlayer electron-electron
distance ($d\approx340$~\AA) is comparable to the lowest used in 
Ref.~\onlinecite{murp95}.
This could suggest that it is the greater disorder 
in our samples that gives rise to
the smaller enhancement for $\Gamma_{\text{e-e}}$.
By way of comparison, we point out that Berk
{\em et al.\/}\cite{berk95} have measured electron-electron
scattering rates similar to our own (in the units used here, $a_2=3$),
in 2DEGs with a similarly low mobility
$\mu\approx50$-$400\times10^3$~cm$^2/$Vs but 
more strongly coupled ($d=180$~\AA). 
The effect of stronger tunneling 
(as quantified by the symmetric-antisymmetric gap $\Delta_{\text{SAS}}$)
on $\Gamma_{\text{e-e}}$ has been 
calculated by Slutzky {\em et al.\/}\cite{slut96}

\end{section}

\begin{section}{Tunneling in a Perpendicular Magnetic Field}
\label{s:perpb}

The application of a perpendicular magnetic field $B$ has a marked 
effect on the resonant tunneling properties between two 2DEGs.
Figure~\ref{f:tunnsdh} shows the equilibrium conductance $G(V_{\text{sd}}=0)$,
at a matched carrier density of $n=1.0\times10^{11}$~cm$^{-2}$,
as a function of magnetic field at 
0.1~K (solid line) and 1.3~K (dashed line).
As $B$ is increased,
the oscillations in the tunneling conductance mirror
the formation of LLs in both 2DEGs.
However, in contrast to SdH oscillations,
which are an in-plane transport effect,
the oscillations shown in Fig.~\ref{f:tunnsdh} 
originate from the current flowing 
perpendicular to the 2DEGs.\cite{note1}
The oscillations in $G(V_{\text{sd}}=0)$
can be interpreted in terms of variations of $\rho_B$,
the DOS at the Fermi level in a field $B$.  
We will describe the temperature dependence of the 
equilibrium tunneling in two magnetic field regimes, 
$B<0.6$~T and $B>0.6$~T.

Below 0.6~T, the formation of LLs modulates the DOS,
creating minima (maxima) at integral (half-integral) filling factors.
The equilibrium tunneling amplitude
depends on the DOS in both layers
and shows quantum oscillations commensurate with $\rho_B$.
When the temperature is increased from 0.1 to 1.3~K,
the low field tunneling conductance 
follows the behavior of $\rho_B$;
the amplitude of the conductance oscillations decrease
as the modulation in $\rho_B$ becomes weaker,
and $G(V_{\text{sd}}=0)$ tends towards its zero-field value.

For $B<0.6$~T the magnitude of the conductance 
maxima increase with magnetic field,
following the expected increase in 
the degeneracy of a half-filled LL.
However, for $B>0.6$~T the amplitude of the
tunneling maxima {\em decrease} with increasing magnetic field;
this can be interpreted as a 
magnetic-field-induced suppression of 
the equilibrium tunneling between the two 2DEGs,
with the formation of a gap in the tunneling DOS.
When the temperature is raised to 1.3~K
the suppression of the conductance is reduced,
indicating thermal activation across the gap.
The Fig.~\ref{f:tunnsdh} inset shows
low-field equilibrium tunneling for a matched carrier density of 
$n = 3\times 10^{11}$~cm$^{-2}$.
At this higher carrier density the crossover from SdH-like to
activated  behavior again occurs at $B=0.6$~T.
These measurements at different temperatures show that
the signature of the magnetic-field-induced gap
is evident at filling factors as high as $\nu \approx 18$.
Clearly, above 0.6~T a mechanism other than the simple
formation of LLs is needed to explain the 
unusual temperature dependence observed 
at half-integral filling factors.

The formation of a magnetic-field-induced gap can also be 
investigated by non-equilibrium measurements,
where at low magnetic fields a dip 
around zero bias is observed in the 
$G(V_{\text{sd}})$ characteristics.
Figure~\ref{f:gap} shows $G(V_{\text{sd}})$ characteristics
at various magnetic fields for a matched
carrier density of $n=0.95\times 10^{11}$~cm$^{-2}$.
In the trace at $B = 0.4$~T
there is a slight dip at $V_{\text{sd}}=0$;
with increasing magnetic field both the depth of the dip 
and the separation of the surrounding peaks become larger.
To characterize the tunneling gap,
we define $\Delta$ to be the voltage separation of the conductance 
peaks either side of $V_{\text{sd}}=0$.
Previously, we have measured\cite{brown94c}
$I(V_{\text{sd}})$ characteristics for $\nu <1 $
and have successfully fitted our low voltage data to the expression 
$I=I_0\exp (-\Delta/eV_{\text{sd}})$.
At high fields the values of $\Delta$ 
obtained from such fits are precisely the same 
as those measured from the peak separations.\cite{brown94e}
Although at low magnetic fields we cannot fit our 
data to this theoretical expression for $I$,
we nevertheless use the peak separation as a 
straightforward measure of the gap.
To test various theoretical predictions 
we have investigated $\Delta$ at fixed $n$ and fixed $\nu$, 
both as a function magnetic field $B$.

Figure~\ref{f:gapb} shows the gap parameter $\Delta$, measured from
traces similar to those in Fig.~\ref{f:gap}, over a wider range of
$B$. Due to the presence of edge states we disregard measurements
close to the integer quantum Hall (QH) regime, as well as those near
the fractional QH state at $\nu=2/3$.  The linear fit through the
remaining points (solid circles) is $\Delta =
(0.45\pm0.02)\hbar\omega_{\text{c}} - (0.19\pm0.05)$~meV.
Figure~\ref{f:gapb} shows that the gap is most cleanly observed at
half-integral filling factors, and in measurements at fixed 
$n$ there are more data points at low $\nu$ than at high $\nu$.  
With this proviso the linear fit in Fig.~\ref{f:gapb} spans
the whole field regime, though it is difficult to identify deviations
at low $B$.  We have also performed conductance measurements
at a higher matched carrier density of $n=1.48\times
10^{11}$~cm$^{-2}$, and Fig.~\ref{f:greyscale} shows the
evolution of the the low field tunneling characteristics 
at this carrier density presented as a grey-scale plot.  
The figure is made up of $G(V_{\text{sd}})$ characteristics,
similar to those in Fig.~\ref{f:gap}, taken at 101 magnetic fields
from $B=0$ to 2~T.  To accentuate the data each trace has been scaled
between 0 (black) and 1 (white).  The light structures around the
center, marked with solid lines, are the two tunneling peaks
surrounding the gap, with a splitting that is linear in magnetic field,
$\Delta=(0.43\pm0.03)\hbar\omega_{\text{c}} - (0.26\pm0.03)$~meV.
Within experimental error the gaps measured at the two carrier
densities, $0.95\times 10^{11}$~cm$^{-2}$ and $1.48\times
10^{11}$~cm$^{-2}$, have the same magnetic field dependence.

To obtain more information about the gap at higher filling factors,
Fig.~\ref{f:gapnu} shows measurements of $\Delta_{\nu}$
taken as a function of $B$ while
maintaining a half-integral filling factor in the tunneling region.
Gap measurements were obtained at $\nu=1/2,\, 3/2,\ldots, 11/2$ for
seventeen matched carrier densities in the range $0.36 \le n \le
3.10\times 10^{11}$~cm$^{-2}$. 
At a given magnetic field, the higher the half-integral 
filling factor, the smaller the gap. 
The DOS is not clearly
spin-split below $B\approx1$~T, and so data points for the highest
filling factor and lowest fields are subject to some error. 
The straight line $0.45\hbar\omega_{\text{c}}-0.19$~meV is included 
in Fig. 11 to demonstrate the linear $B$ dependence of the tunneling 
gap at $\nu=1/2$, and the sublinear nature 
of the gap at higher filling factors.
At low carrier densities $n<1.5\times10^{11}$~cm$^{-2}$, measurements 
of $\Delta_{\nu\geq3/2}$ converge to a value
similar to that measured at $\nu=1/2$.
For the higher carrier concentrations the gap for $\nu>3/2$ is 
reduced in comparison with $\Delta_{1/2}$.
This observation, plus the scarcity of
data points at higher filling factors in measurements at fixed $n$
accounts for the same measured gap at $0.95\times
10^{11}$~cm$^{-2}$ and $1.48\times 10^{11}$~cm$^{-2}$.
We stress that, as previously reported,\cite{brown94c} the high field
gap shows the same linear behavior both at 
fixed $n$ and fixed $\nu<1$.

Figures~\ref{f:gap} and~\ref{f:greyscale} also show tunneling between
LLs of different index.  At high magnetic fields, the first inter-LL
tunneling transition has been observed\cite{brown94c,eis92a} as a peak
in the $I(V_{\text{sd}})$ characteristics at $eV_{\text{sd}}\approx
1.3 \hbar\omega_{\text{c}}$, and the equivalent conductance
measurement\cite{brown94e} exhibits a peak at $eV_{\text{sd}} \approx
1.16 \hbar\omega_{\text{c}}$.  Both measurements show inter-LL
transitions occuring at higher than expected bias voltages, and it is
proposed\cite{eis92a} that this enhanced LL spacing is due to an
effective mass reduced by many-body effects.\cite{smith92} The arrows
in Fig.~\ref{f:gap} and the dashed lines superimposed on
Fig.~\ref{f:greyscale} show the position of the first inter-LL
tunneling peaks either side of zero bias. Over the range
$0.5\le B\le 3$~T the conductance peaks occur at
$eV_{\text{sd}}=(1.07 \pm 0.03)\hbar\omega_{\text{c}}$, an enhancement
that is only slightly less than that observed at higher magnetic
fields.  If the enhanced LL spacing is due to a reduced effective
mass, our measurements suggest that the mass is reduced at higher
filling factors by an amount much greater than that predicted by Smith
{\em et al.\/}\cite{smith92}

\subsection*{Comparison with Theory}\label{s:comp}

There has been much interest in tunneling in a weak magnetic field
where, even for filling factors $\nu \gg 1$,
it is believed that there is an energy cost associated with the 
injection and extraction of electrons into and out of a 2DEG.
It has been theoretically shown\cite{aleiner95a} that at 
low magnetic fields the bare Coulomb interaction 
(which typically exceeds the cyclotron energy $\hbar \omega_{\text{c}}$)
is screened by the filled LLs,
and the renormalized interaction is sufficiently 
weak that it does not mix LLs.
Therefore only interactions between electrons in the 
outer partially filled LL need to be considered;
this is the starting point for many of the theories
in the weak field regime.

Following the treatment at high magnetic fields, 
the gap at lower fields and higher filling factors
can be considered to be the Coulomb energy,
where the cyclotron length $l_{\text{c}}=\hbar k_{\text{F}}/eB$ 
replaces the magnetic length $l_B$ as the relevant length scale.
The semiclassical energy gap is
\begin{equation}
\Delta_1=\frac{e^2}{4\pi\epsilon l_{\text{c}}} =
\frac{r_{\text{s}}\hbar\omega_{\text{c}}}{\sqrt{2}},
\label{e:semigap}
\end{equation}
where $r_{\text{s}}$ is the interaction parameter,
and is given by 
$r_{\text{s}} = a/a_{\text{B}} = \sqrt{2}/a_{\text{B}}k_{\text{F}}$.
Levitov and Shytov\cite{levi95b} consider the problem
of an electron orbiting in a ring of thickness 
$l_B$ and radius $l_c$ in one layer,
tunneling into a similar ring in the second layer, 
leaving behind a hole in the first ring.
The electrostatic energy of this two-ring capacitor
creates an energy gap that is proportional to Eq.~\ref{e:semigap},
but multiplied by a logarithmic screening term.
Alternatively, using both a hydrodynamic approach\cite{aleiner95a}
and a Hartree-Fock calculation,\cite{aleiner95b}
Aleiner and coworkers have obtained a gap of the form 
\begin{equation}
\Delta_2 = \frac{\hbar\omega_{\text{c}}}{2\nu} \ln(r_{\text{s}}\nu).
\label{e:glazman}
\end{equation}
It has been shown\cite{levi95b} that
the total gap has contributions from two terms similar
in form to $\Delta_1$ and $\Delta_2$, 
where $\Delta_1$ is the dominant term.

In a recent theory, Fogler {\it et al.}\cite{fogler96} propose that
under certain conditions the ground state of the 2DEG in a weak
magnetic field is a charge density wave (CDW) superimposed on a
uniform background of filled LLs.  Tunneling into a partially filled
LL disturbs some of the correlations of the CDW, and the energy cost
of this disruption gives rise to a tunneling gap
\begin{equation}\label{e:fog}
\Delta_3 = \frac{r_{\text{s}} \hbar\omega_{\text{c}}}{\sqrt{2} \pi}
\ln\left(1+\frac{0.3}{r_{\text{s}}}\right) + \Delta_2 \approx 
0.07\hbar\omega_{\text{c}}+\Delta_2.
\end{equation}
As $B$ is decreased there is a predicted\cite{fogler96} crossover from
the spin-resolved QH regime (spin-split LLs) where the ground state is
a CDW, to the spin-unresolved QH regime (well separated, but
spin-degenerate LLs).  The CDW ground state can only exist in the
former regime, while at lower fields the ground state is a uniform
electron liquid which is expected\cite{fogler96} to have a gap of a
different origin.  In our samples the SdH oscillations of the
individual 2DEGs show that the spin-unresolved QH regime occurs over a
narrow range of magnetic field.  The Fig.~\ref{f:tunnsdh} inset shows
that at $n=3\times 10^{11}$~cm$^{-2}$ it is possible to observe
a suppression of tunneling (above 0.6~T) in the absence of
spin-splitting (which becomes distinct for $B>1.5$~T).  Moreover, in
all our measurements over a wide range of $n$, we see no clear change
in the character of the gap once spin-splitting has set in.  The
$B$ dependence of the low field gaps predicted by Aleiner {\it et
al.}\cite{aleiner95b} and Fogler {\it et al.}\cite{fogler96} are
determined by $\Delta_2$, which varies as $\Delta_2 \sim B^2$ at fixed
carrier density, and as $\Delta_2 \sim B$ at fixed filling factor.
Figure~\ref{f:gapb} shows that at fixed $n$ the gap is linear (not
quadratic) in $B$, and Fig.~\ref{f:gapnu} shows that
at fixed $\nu$ the weak field
gap $\Delta_{\nu \geq 3/2}$ is sublinear in $B$.

The semiclassical gap (Eq.~\ref{e:semigap})
behaves as $\Delta_1 \sim B$ at fixed
carrier concentration, and as $\Delta_1 \sim \sqrt{B}$ at fixed filling
factor.  For fixed $n$ the measured gaps are indeed
linear in $B$, though we find that the best fits are obtained with a
negative intercept.  Similarly the best fits of $\Delta_{\nu}$ to a
$\sqrt{B}$ field dependence are also 
obtained with a negative intercept, 
using the expression $\Delta_\nu =a'\sqrt{B/\nu}-b'$. 
In the inset to Fig.~\ref{f:gapnu} we have plotted 
$(\Delta_{\nu}+b')\nu^{1/2}$ versus $B^{1/2}$ for $\nu \geq 5/2$.  
Taking $b'=0.4$~meV the data collapse onto a single
straight line with slope $a'=2.0$.  
We conclude from measurements at
constant $n$ and constant $\nu$ that the high filling factor gap is
best described by
\begin{equation}
\label{e:modgap}
\Delta = (0.46\pm0.03) \frac{e^2}{4\pi\epsilon l_{\text{c}}} 
-(0.4\pm0.1)\text{~meV}.
\end{equation}
The negative intercept could be interpreted as 
evidence for an excitonic effect
(similar to that observed\cite{eis95} at $\nu=1/2$),
and with a magnitude which is approximately 
independent of filling factor for $\nu \geq 5/2$.
We were unable to obtain good agreement between Eq.~\ref{e:modgap}
and the measured gaps at $\nu=1/2$ and 3/2,
instead much larger negative intercepts 
($b'=4.2$ and 1.7~meV, respectively) were required.
With our samples we have obtained\cite{brown94e} better agreement at $\nu=1/2$
using a linear rather square root magnetic field dependence
for the gap $\Delta_{1/2}$ up to 25~T.

\end{section}

\begin{section}{Temperature Studies in High Fields}
\label{s:temp}

In this section we use Eq.~\ref{e:spect} to model 
high field $G(V_{\text{sd}})$ measurements taken at $\nu=1/2$,
results which have been presented elsewhere.\cite{turn96a}
Figure~\ref{f:tempdep}(a) shows the conductance characteristics
at $B=8$~T at a matched carrier density of 
$n_1=n_2=0.97 \times 10^{11}$~cm$^{-2}$.
As the temperature is increased from $T=1.5$~K to 6~K,
the equilibrium conductance $G(V_{\text{sd}}= 0)$ increases,
while the height of the surrounding peaks is reduced.
Figure~\ref{f:tempdep}(b) shows $G(V_{\text{sd}}=0)$ in an Arrhenius plot, 
revealing activated behavior at low temperatures
with an activation energy of $E_{\text{a}}= 0.35 \pm 0.01$~meV.
Above the activation temperature ($\approx 4$~K)
the equilibrium conductance departs from exponential behavior,
and for $T>10$~K the conductance decreases. 
We have measured $E_{\text{a}}$ at different 
matched carrier densities and magnetic fields,
while maintaining a filling factor of $\nu=1/2$ in the tunneling region.
Figure~\ref{f:eab} shows $E_{\text{a}}$ 
as a function of magnetic field, 
with a fit which shows that the activation 
energy is proportional to $B$.
At $\nu = 1/2$ the ratio $E_{\text{a}}/\Delta$ is $0.047\pm0.005$,
and is approximately independent of $B$.
Activation plots of the high field gap
were first obtained by Eisenstein {\em et al.\/},
and for comparison we measure 
from Fig.~3 of Ref.~\onlinecite{eis92a} the ratio
$E_{\text{a}}/\Delta = 0.07$ at $B=13$~T.

Temperature can be used to probe the shape of the tunneling DOS,
and previously Ashoori {\em et al.\/}\cite{ash93} have used
a DOS with a linear dip at the Fermi level 
to model 2D-3D tunneling measurements.
A linear DOS does not give activated behavior,
and to obtain such characteristics 
we have modeled the data using 
the double-Gaussian spectral density
\begin{equation}
\label{e:alsdos}
A(E) \propto \frac{1}{\sqrt{E_0 k_{\text{B}}T}}\left\{
\exp\left(-\frac{(E_0-E)^2}{2\alpha E_0 k_{\text{B}}T}\right)+
\exp\left(-\frac{(E_0+E)^2}{2\alpha E_0 k_{\text{B}}T}\right)\right\},
\end{equation}
where the Fermi level $E_{\text{F}}$ lies in the minimum between the
two peaks.  This tunneling DOS is a modified version of Eq.~14 in
Ref.~\onlinecite{aleiner95a}, which was originally proposed for
$\nu\gg 1$ and $E_0 \gg k_{\text{B}}T$.  By comparison with the
original formula, we have introduced an adjustable parameter $\alpha$
in the denominator of the exponential.  As $T \rightarrow 0$ the
double-Gaussian spectral density consists of two $\delta$-functions at 
$E =\pm E_0$, and so the gap $\Delta \rightarrow 4E_0$ at $T=0$.  At low
temperatures the activation energy $E_{\text{a}} \rightarrow
E_0/\alpha$, and so $\Delta$ and $E_{\text{a}}$ share the same
magnetic field dependence (that of $E_0$), in agreement with
experiment.

To simulate the conductance measurements in Fig.~\ref{f:tempdep}
we have chosen $\alpha=5.3$ so that the theoretical
ratio $E_{\text{a}}/\Delta = 1/4\alpha$
agrees with the experimental value of 0.047.
$I(V_{\text{sd}})$ characteristics have been calculated 
using Eq.~\ref{e:alsdos} in Eq.~\ref{e:spect},
and by numerical differentiation of these characteristics 
we obtain the $G(V_{\text{sd}})$ traces shown in Fig.~\ref{f:gausst}(a).
The value of $E_0=2.6$~meV best reproduces the conductance 
and activation characteristics (see Fig.~\ref{f:gausst}(b)) at 8~T.
From similar simulations of $\nu=1/2$ data at different magnetic fields
we have measured the $B$ dependence of $E_0$;
the results are displayed in Fig.~\ref{f:eab},
with a fit showing $E_0$ with a linear dependence on 
the magnetic field, $E_0 =0.2\hbar\omega_{\text{c}}$.

The calculated $G(V_{\text{sd}})$ curves reproduce many features of
the experimental data.  First, there are bias voltages which delineate
regions of positive and negative $\partial G/\partial T$; second, the
conductance maxima move closer together with increasing temperature;
third, it shows that at higher temperatures $G(V_{\text{sd}}=0)$ is no
longer activated, but reduces with increasing temperature.  There are,
however, some differences between the model and experiment.  The very
low temperature limit of the experimental characteristics differ from
those of the model; in the experimental system disorder introduces a
finite width to the DOS at $T=0$.  We have tried to simulate this
disorder by introducing a finite offset temperature, that is by
replacing $T$ with $T + T_0$ in Eq.~\ref{e:alsdos}, where $T_0$ is a constant.
However, there was little improvement in the shape of the simulated
characteristics.

\end{section}

\begin{section}{Conclusions}
\label{s:conclude}

We have used both the equilibrium and non-equilibrium tunneling characteristics
to show that the underlying spectral function, $A(\bbox{k},E)$, within
each of the 2DEGs of a double-layer system is a Lorentzian.
Equilibrium measurements show that the Lorentzian tunneling linewidth has a
carrier density and temperature dependence that can be attributed to
electron-impurity and electron-electron scattering.  The measured
electron-impurity scattering rate is similar in magnitude to the
Dingle time, and comparison with theory shows that correlations
between scatterers need to be taken into account to obtain better
agreement with experiment.  The mathematical form,
but not the magnitude, of the electron-electron scattering rate
measured from the equilibrium lineshape is well described by theory.

In a perpendicular magnetic field the equilibrium tunneling
conductance oscillates as a function of $B$, reflecting the formation
of LLs.  For $B> 0.6$~T the magnetic field creates a gap in the
tunneling DOS which suppresses the equilibrium tunneling.  For fixed
filling factor $v=1/2$ the tunneling gap has a linear magnetic field
dependence, whereas at higher half-integral filling factors the gap is
sublinear and is best described by a semiclassical gap.  We have
modeled the temperature dependence of the conductance characteristics
in a strong magnetic field using a double-Gaussian spectral density.

\section*{Acknowledgments}

We wish to thank the Engineering and Physical Sciences 
Research Council (UK) for supporting this work. 
JTN acknowledges support from the Isaac Newton Trust,
and DAR acknowledges support from Toshiba Cambridge Research Centre.   
We thank A.V. Khaetskii for a critical reading of the manuscript,
and we acknowledge help with the numerical calculations
from I. M. Castleton and C. H. W. Barnes.

\end{section}


\begin{figure}
\caption{Conduction band profile of the double 2DEG structure,
defining the various energies used in Sec.~\ref{s:formal}.}
\label{f:band}
\end{figure}

\begin{figure}
\caption{Equilibrium tunneling conductance 
$G(V_{\text{sd}}=0)$ as a function of the gate 
voltage $V_{\text{g}_1}$ controlling the carrier
density in the upper layer, 
when the lower layer density is fixed at
$n_2=3.25\times 10^{11}$~cm$^{-2}$.
The traces were taken at $T=3$~K (squares) and 19~K (circles),
and the solid lines are the fits to Eq.~\ref{e:lorentz2}.}
\label{f:lorentz}
\end{figure}

\begin{figure}
\caption{Comparison of non-equilibrium (squares) 
and equilibrium (circles) tunneling lineshapes at 0.1~K.
The solid lines are fits to Eqs.~\ref{e:noneqbmgen}
and~\ref{e:lorentz2}, respectively.
For clarity not all the experimental data points are shown, and the
equilibrium data and fit have been vertically offset by 1~$\mu$S.}
\label{f:betafig}
\end{figure}

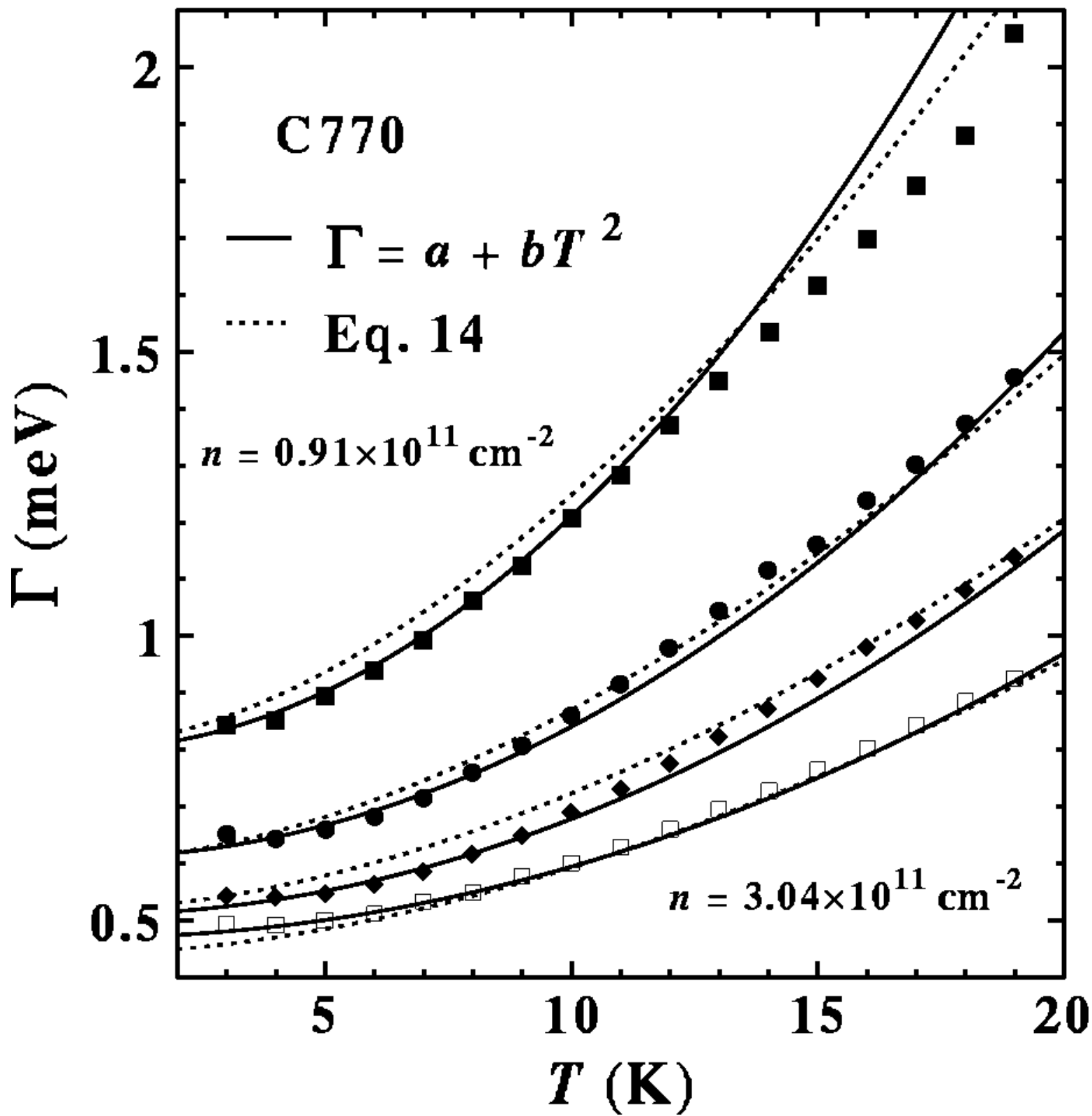
\begin{figure}
\caption{Temperature dependence of the equilibrium linewidth $\Gamma$ for 
$n_1=n_2= 0.91,\, 1.62,\,  2.19,\, 3.04 \times 10^{11}$~cm$^{-2}$.
The solid lines are fits to the form $a+bT^2$,
and the dashed lines are obtained from the best fit to Eq.~\ref{e:ratent}.}
\label{f:widtht}
\end{figure}

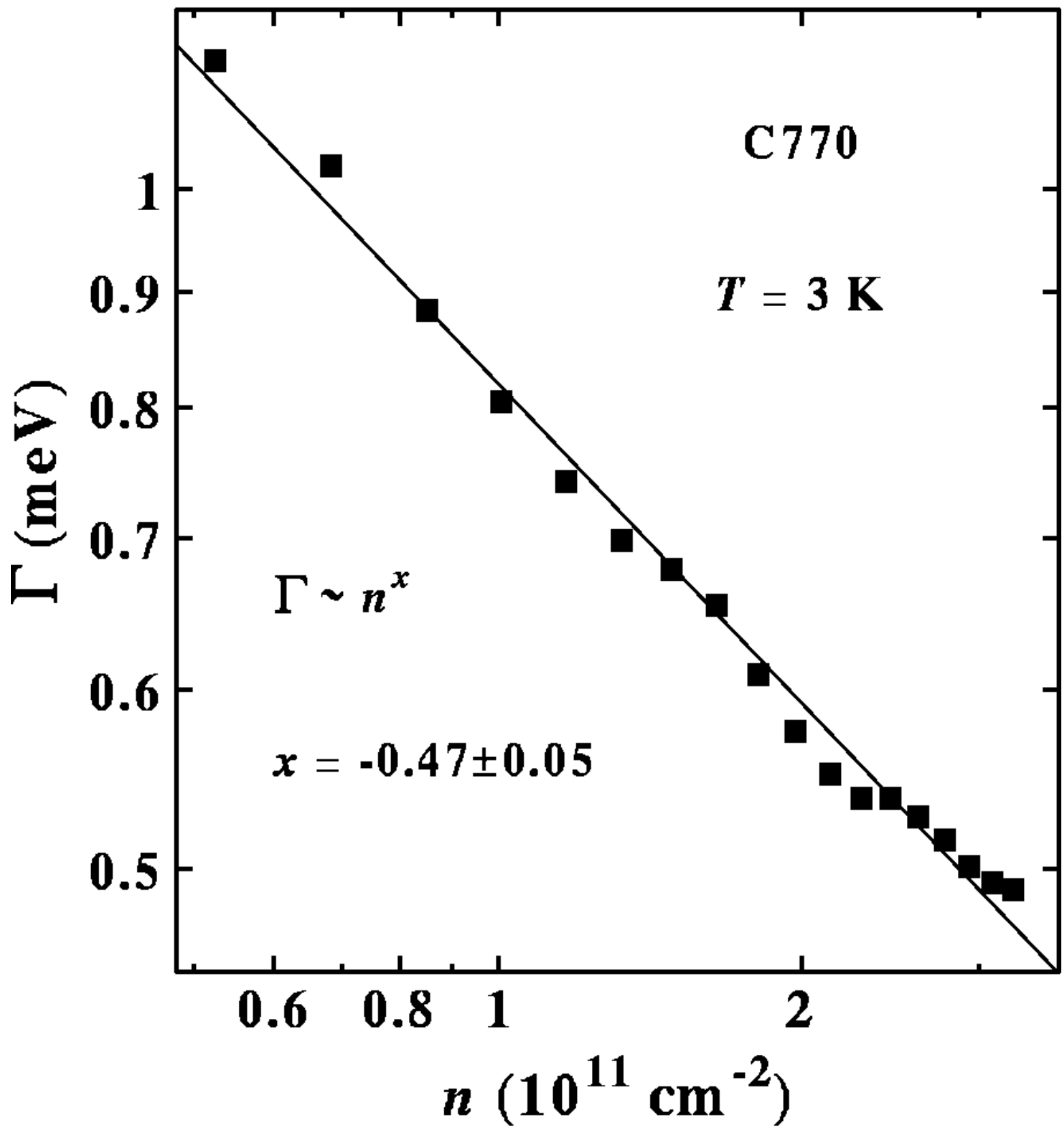
\begin{figure}
\caption{Log-log plot of the linewidth $\Gamma$ versus carrier
density $n$ at $T=3$~K.}
\label{f:logglogn}
\end{figure}

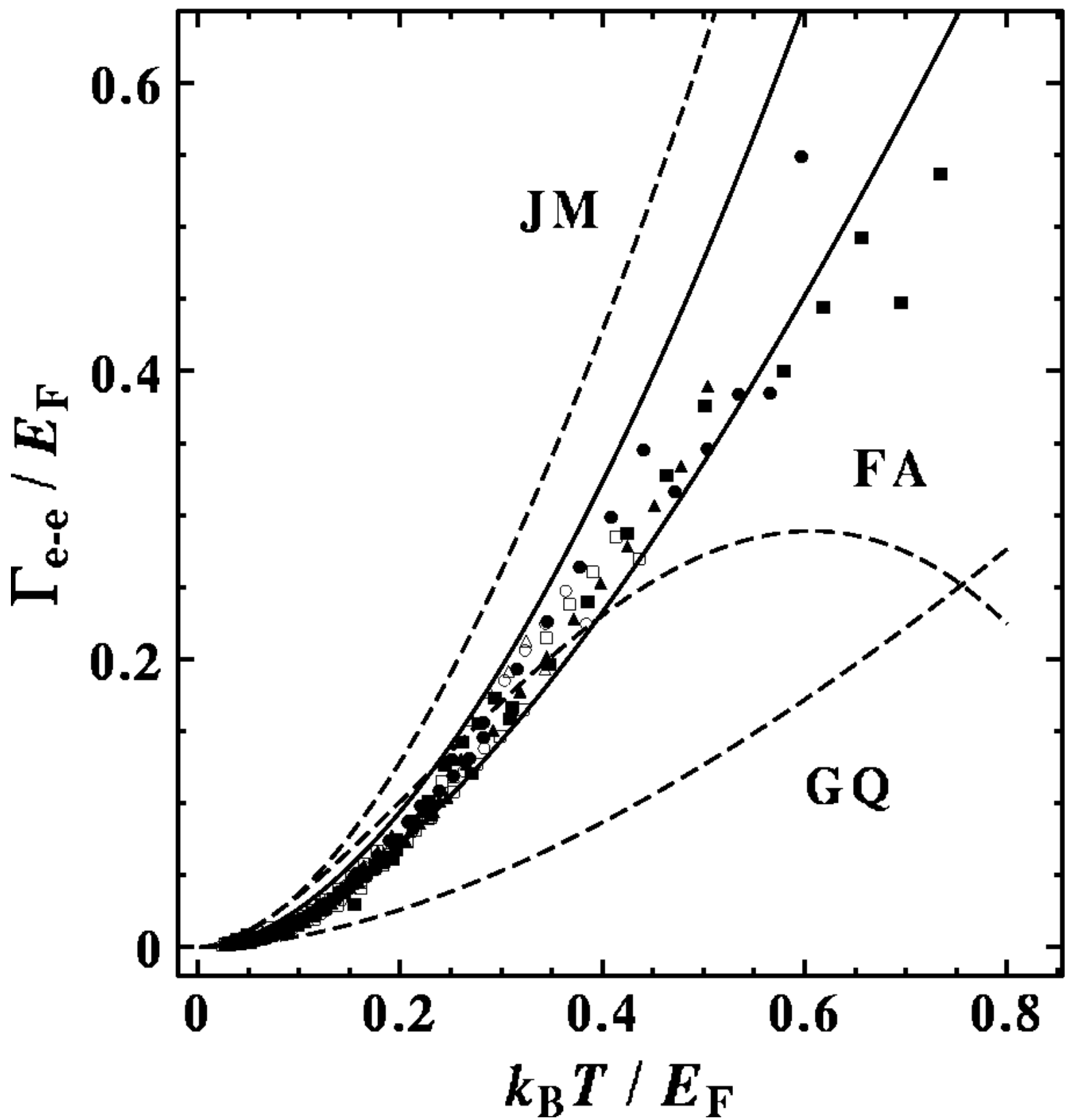
\begin{figure}
\caption{$\Gamma_{\text{e-e}}/E_{\text{F}}$ plotted versus
$k_{\text{B}}T/E_{\text{F}}$. The dashed curves are the 
theoretical predictions of GQ,\cite{giuliani82} FA,\cite{fuk83}
and JM\cite{jung96a} plotted for $n=1.3\times10^{11}$~cm$^{-2}$.
The solid curves are the best fit to Eq.~\ref{e:gnq} 
with a prefactor of $a_2=3.06$,
and with $n=0.3$ and $3\times10^{11}$~cm$^{-2}$ 
(upper and lower curves, respectively).}
\label{f:univ}
\end{figure}

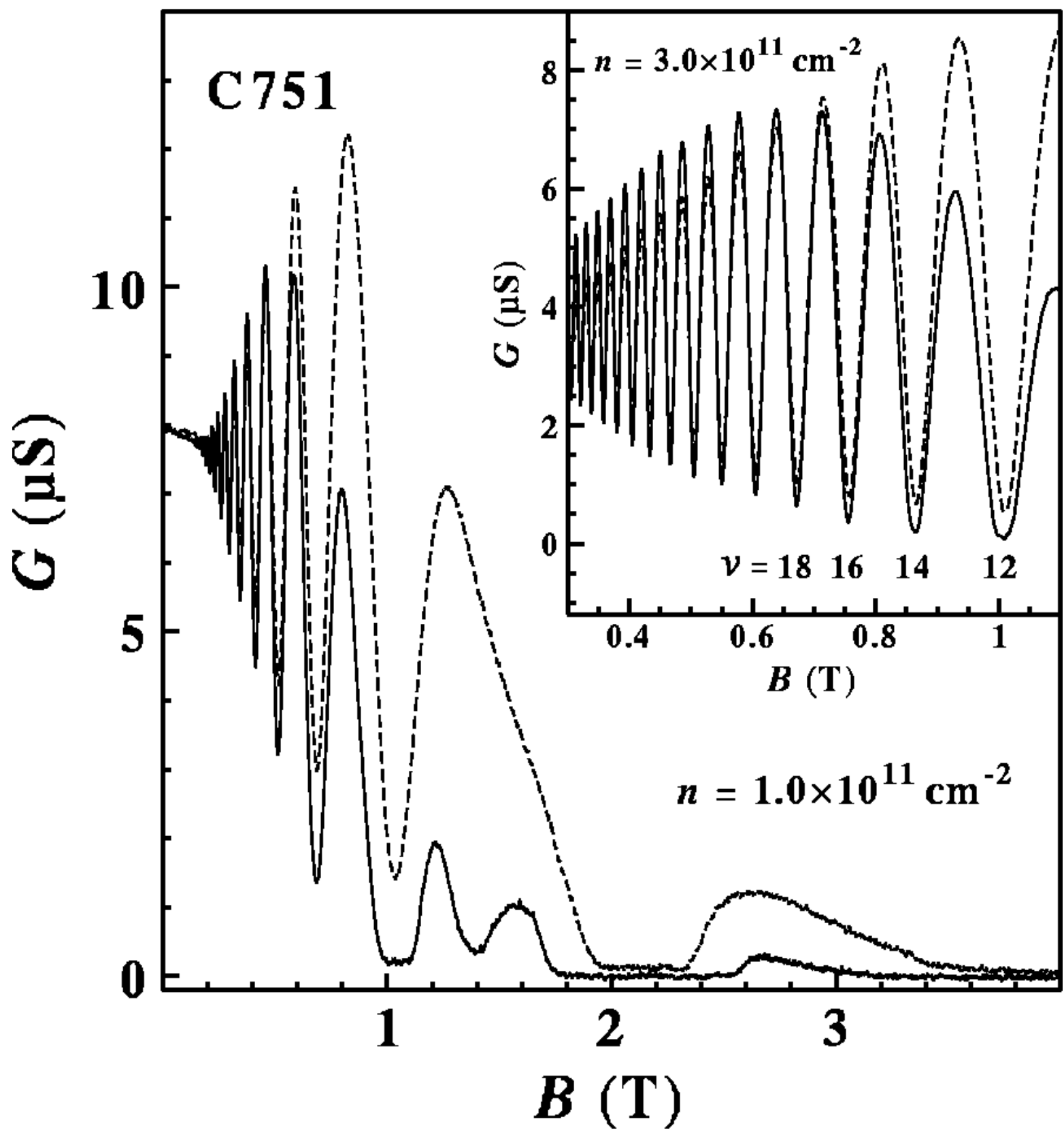
\begin{figure}
\caption{Main figure: equilibrium tunneling conductance 
at a matched carrier density of $n=1.0\times10^{11}$~cm$^{-2}$,
taken as a function of perpendicular magnetic field.
The solid line shows data taken at $T=0.1$~K,
and the dashed line shows data taken at $T=1.3$~K.
Inset: similar measurements taken 
at $n=3.0\times10^{11}$~cm$^{-2}$.
Filling factors for four of the minima are indicated.}
\label{f:tunnsdh}
\end{figure}

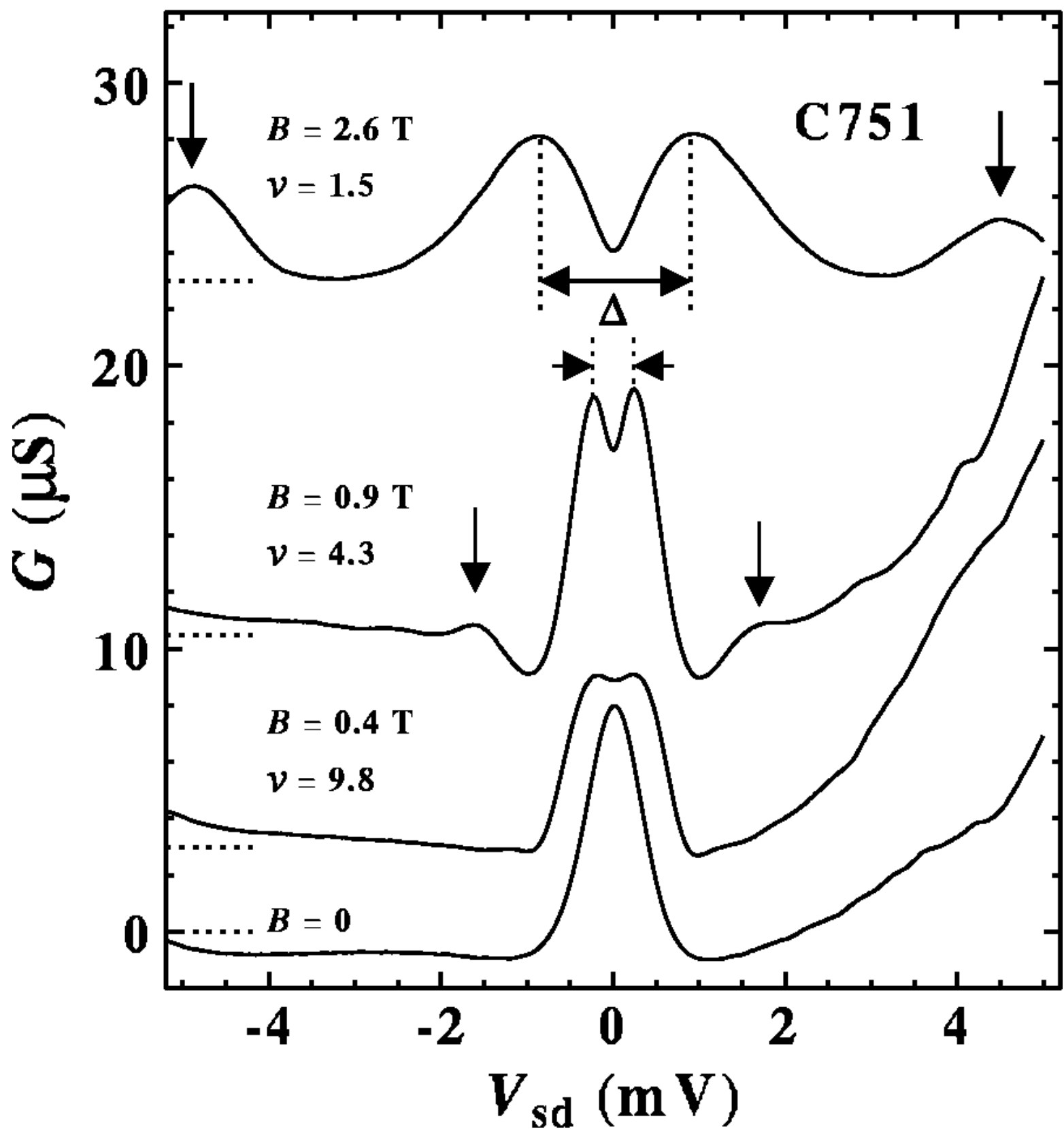
\begin{figure}
\caption{The conductance characteristics $G(V_{\text{sd}})$
in a perpendicular magnetic field at $T=0.1$~K,
taken at a matched carrier density of $n=0.95\times 10^{11}$~cm$^{-2}$.
Sweeps were taken at $B=0,\, 0.4,\, 0.9,\, 2.6$~T 
with the filling factors indicated. 
The curves are offset vertically with 
zeros indicated by the dashed lines.
The splitting of the zero field resonance into two peaks 
defines the gap of width $\Delta$.
Arrows indicate the position of inter-Landau level transitions.}
\label{f:gap}
\end{figure}

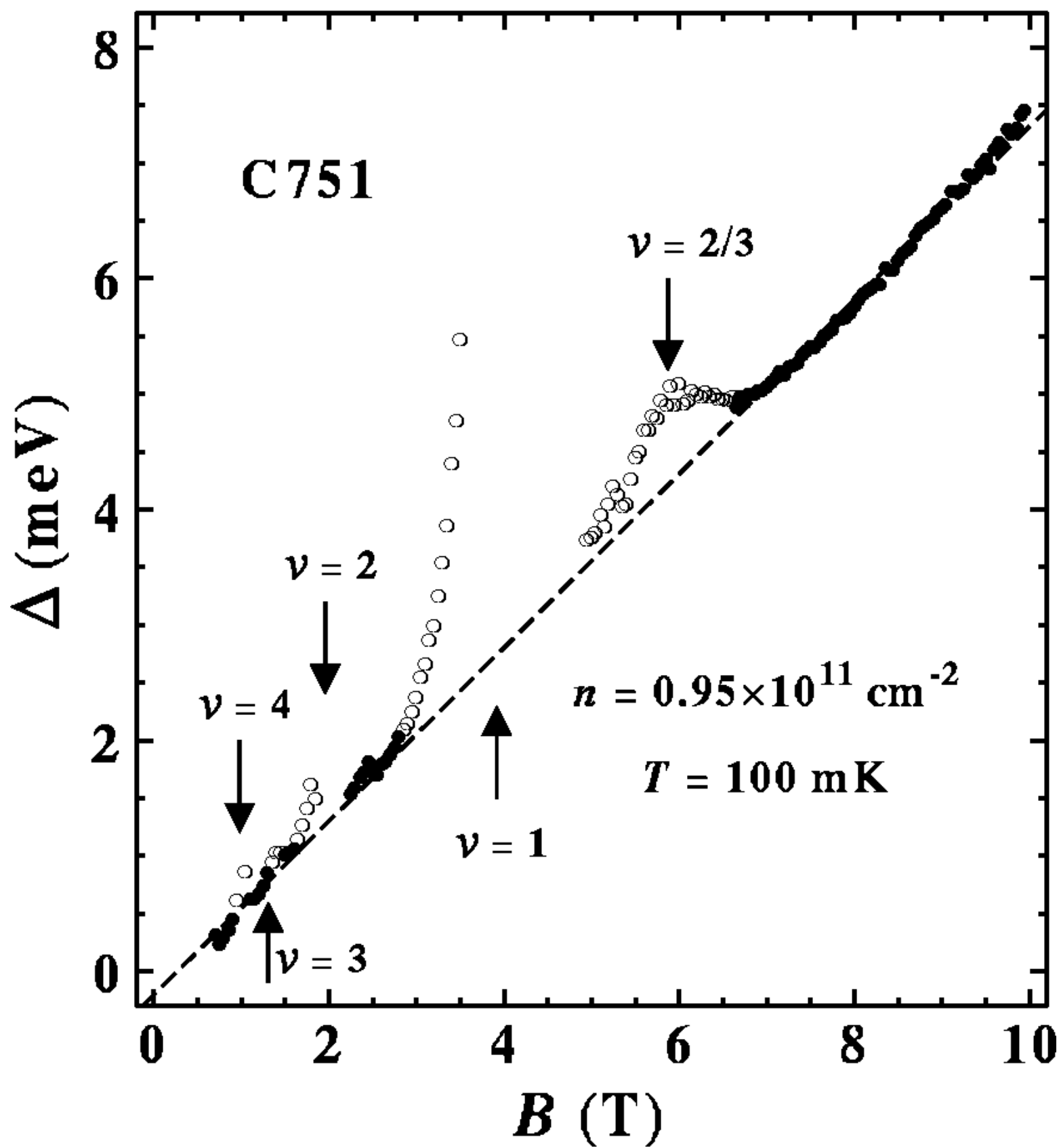
\begin{figure}
\caption{The gap parameter $\Delta$ at matched carrier density
$n = 0.95\times 10^{11}$~cm$^{-2}$ as a function of magnetic field,
measured from traces similar to those shown in Fig.~\ref{f:gap}.
The linear fit, $\Delta = 0.45 \hbar \omega_{\text{c}} - 0.19$~meV,
disregards measurements (open circles) where the 
filling factor approaches integral or fractional (2/3) values.}
\label{f:gapb}
\end{figure}

\begin{figure}
\caption{A grey-scale plot of the tunneling conductance $G$
as a function of $V_{\text{sd}}$ and magnetic field $B$ for 
$n_1=n_2=1.48\times 10^{11}$cm$^{-2}$.
Light areas correspond to regions of high conductance.
The solid lines show the splitting of the tunneling resonance,
with a fitted separation of $\Delta = 0.43 \hbar \omega_{\text{c}} - 0.26$~meV.
Dashed lines indicate inter-LL transitions at
$eV_{\text{sd}} = \pm 1.07 \hbar \omega_{\text{c}}$.
The horizontal stripes at $B=1$~T and 1.5~T
show where the tunneling is completely suppressed in the QH regime
at filling factors $\nu=6$ and $\nu=4$, respectively.}
\label{f:greyscale}
\end{figure}

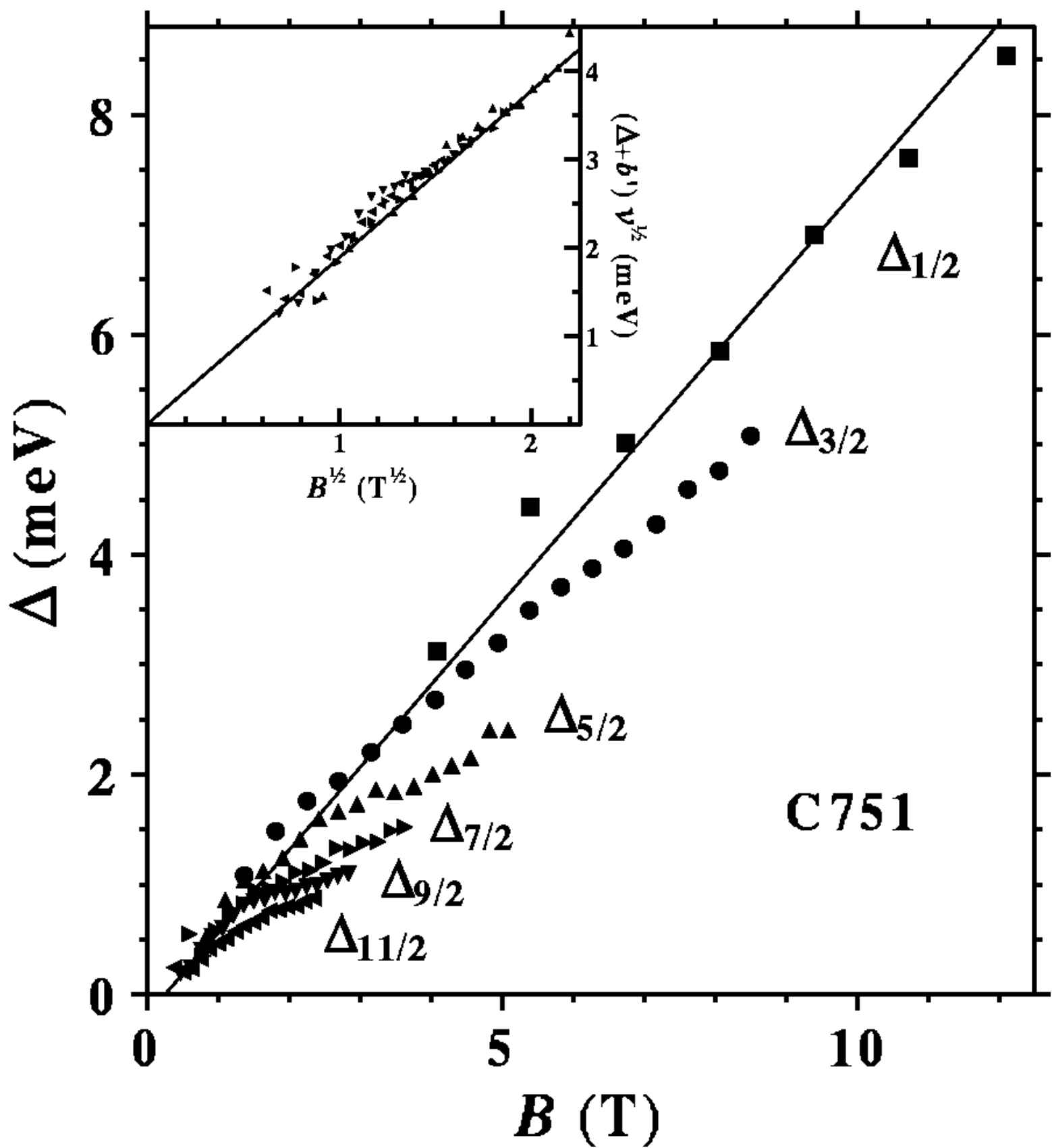
\begin{figure}
\caption{Main figure: the gap parameter $\Delta_{\nu}$ at fixed half-integral 
filling factors $\nu$ as a function of magnetic field.
The solid line shows the expression
$\Delta_{1/2} = 0.45 \hbar \omega_{\text{c}} - 0.19$~meV.
Inset: $(\Delta_{\nu}+0.4\text{~meV})\nu^{1/2}$ plotted versus $B^{1/2}$ for 
$\nu \geq 5/2$. The slope of the solid line gives $a'=2.0$.}
\label{f:gapnu}
\end{figure}

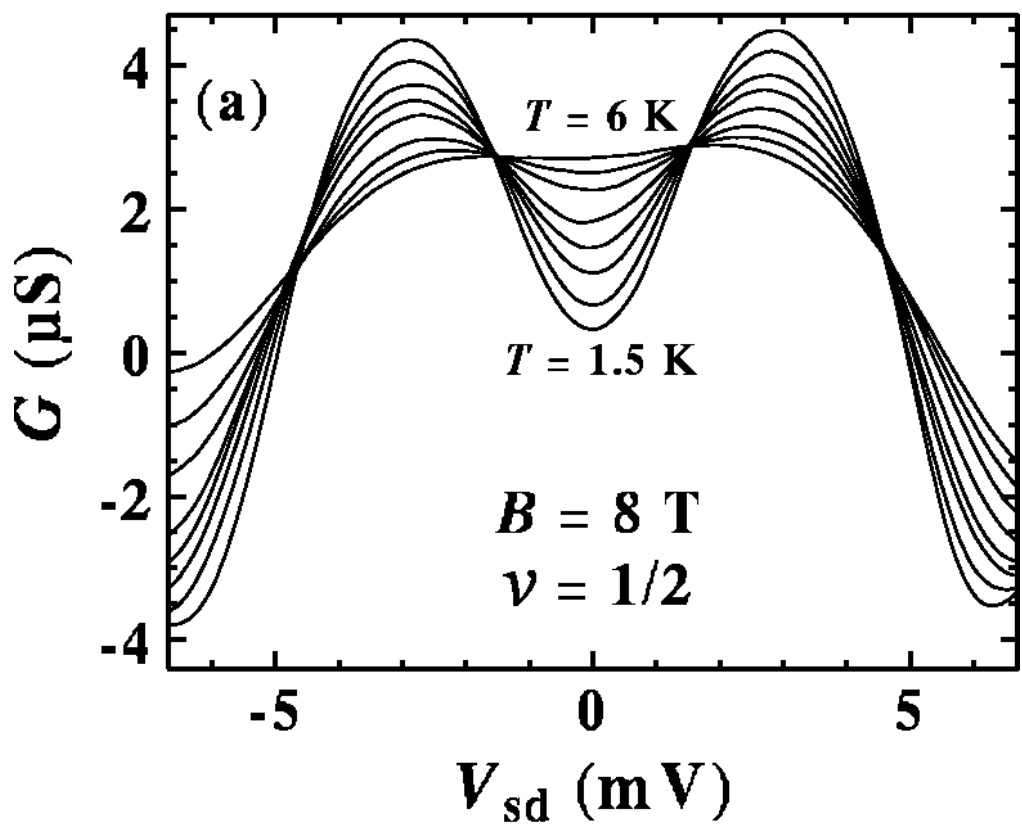
\begin{figure}
\caption{(a) The conductance characteristics $G(V_{\text{sd}})$ 
at $B=8$~T and $\nu = 1/2$ at temperatures of
$T=1.5,\, 1.9,\, 2.4,\, 2.8,\, 3.5,\, 4.3,\, 5.0,\, 6.0$~K.
(b) Arrhenius plot of $G(V_{\text{sd}}=0)$ versus $1/T$ at
$B=8$~T and $\nu=1/2$.
The linear fit at low temperatures gives an activation
energy $E_{\text a} = 0.35\pm 0.01$~meV.}
\label{f:tempdep}
\end{figure}

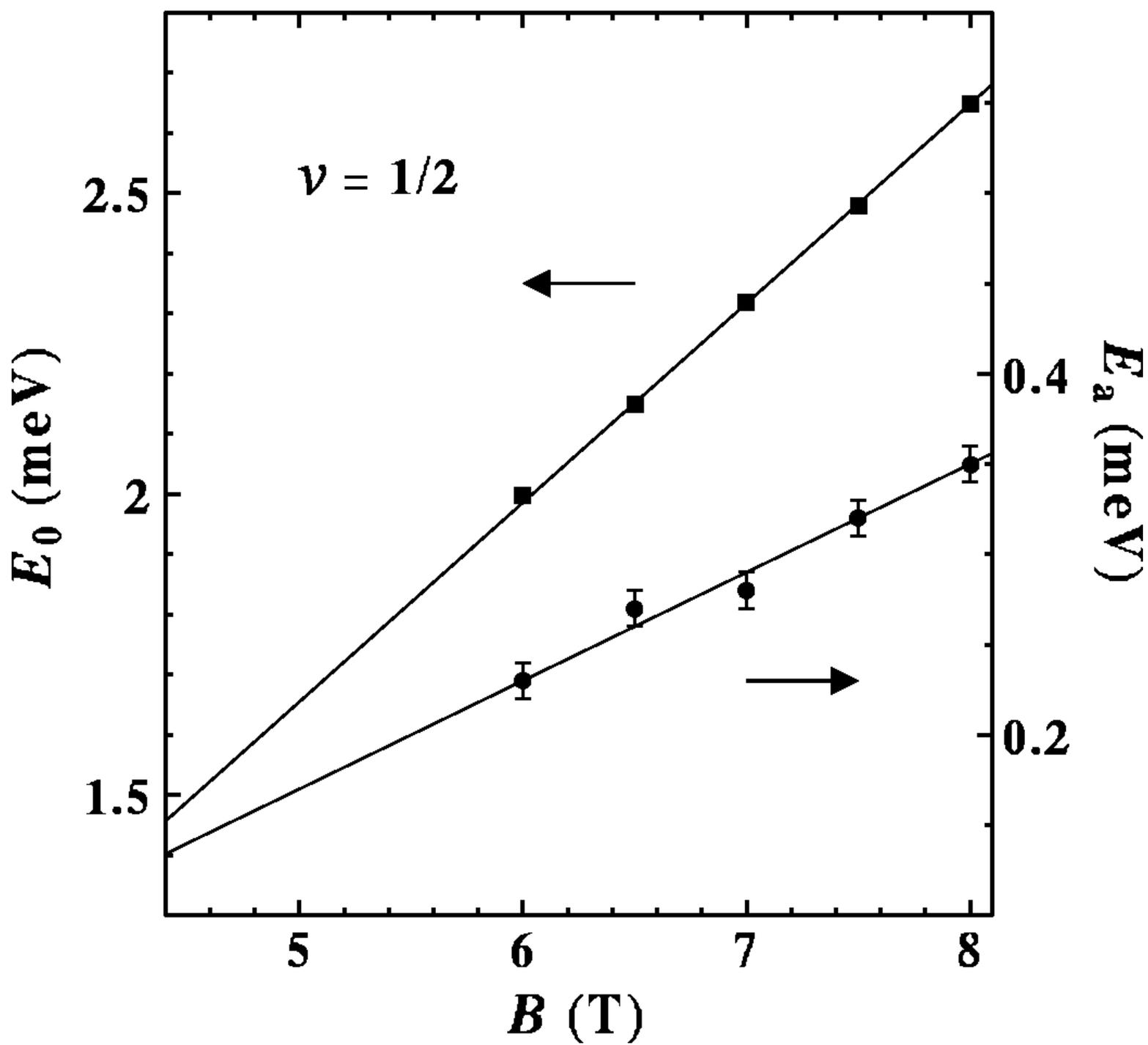
\begin{figure}
\caption{The measured activation energy $E_{\text a}$ at $\nu=1/2$
(circles) and the best fit value of $E_0$ (squares),
as a function of magnetic field.}
\label{f:eab}
\end{figure}

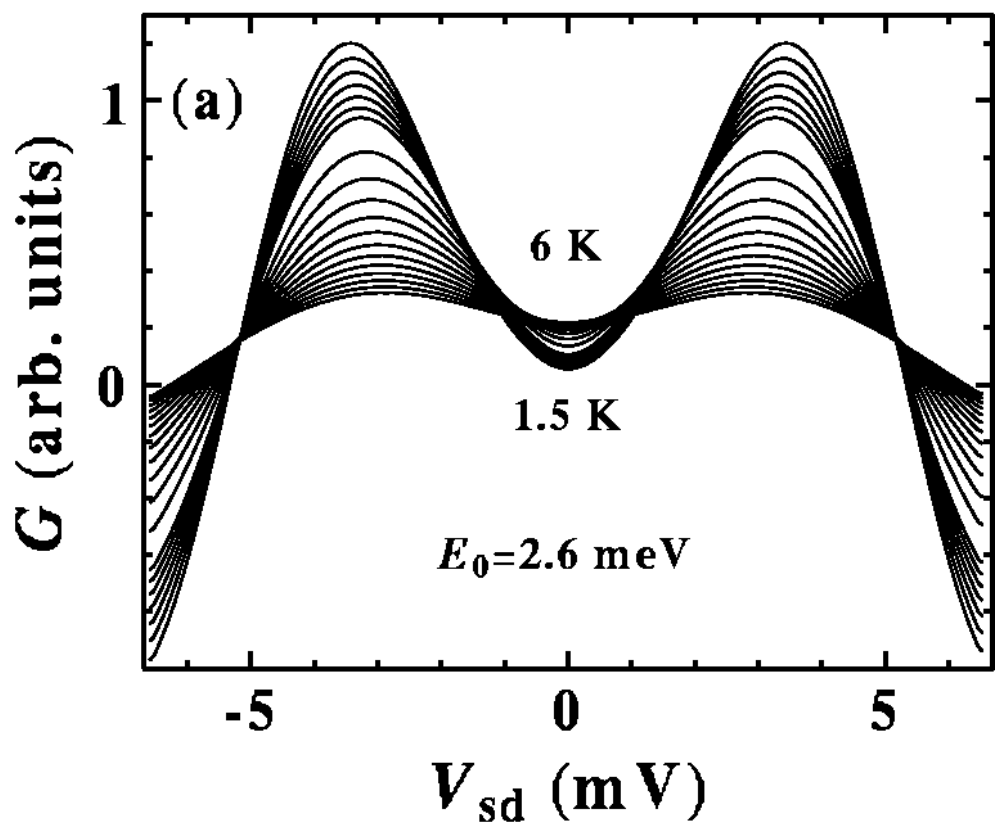
\begin{figure}
\caption{(a) Conductance curves calculated using the
double-Gaussian spectral density (Eq.~\ref{e:alsdos})
with $E_0 = 2.6$~meV and $\alpha=5.3$.
(b) Arrhenius plot of the calculated conductance $G(V_{\text{sd}}=0)$, 
showing an activation energy of $0.35$~meV.}
\label{f:gausst}
\end{figure}


\begin{references}

\bibitem{smol89}
J. Smoliner {\it et~al.}, Phys.\ Rev.\ Lett. {\bf 63},  2116  (1989).

\bibitem{murp95}
S.~Q. Murphy, J.~P. Eisenstein, L.~N. Pfeiffer, and K.~W. West, Phys.\ Rev.\ B
  {\bf 52},  14825  (1995).

\bibitem{yang93}
S.-R.~E. Yang and A.~H. MacDonald, Phys.\ Rev.\ Lett. {\bf 70},  4110  (1993).

\bibitem{hats93}
Y. Hatsugai, P.-A. Bares, and X.~G. Wen, Phys.\ Rev.\ Lett. {\bf 71},  424
  (1993).

\bibitem{song93a}
S. He, P.~M. Platzman, and B.~I. Halperin, Phys.\ Rev.\ Lett. {\bf 71},  777
  (1993).

\bibitem{johan93}
P. Johansson and J.~M. Kinaret, Phys.\ Rev.\ Lett. {\bf 71},  1435  (1993).

\bibitem{brown94c}
K.~M. Brown {\it et~al.}, Phys.\ Rev.\ B {\bf 50},  15465  (1994).

\bibitem{brown94e}
K.~M. Brown {\it et~al.}, Physica B {\bf 211},  430  (1995).

\bibitem{eis95}
J.~P. Eisenstein, L.~N. Pfeiffer, and K.~W. West, Phys.\ Rev.\ Lett. {\bf 74},
  1419  (1995).

\bibitem{ash93}
R.~C. Ashoori, J.~A. Lebens, N.~P. Bigelow, and R.~H. Silsbee, Phys.\ Rev.\ B
  {\bf 48},  4616  (1993).

\bibitem{ashpriv}
R. C. Ashoori, private communication, (1995).

\bibitem{brown94d}
K.~M. Brown {\it et~al.},  in {\em Proceedings of the 22nd Conference on the
  Physics of Semiconductors, Vancouver, Canada, 1994}, edited by D.~J. Lockwood
  (World Scientific, Singapore, 1995), p.\ 1035.

\bibitem{brown94a}
K.~M. Brown {\it et~al.}, Appl.\ Phys.\ Lett. {\bf 64},  1827  (1994).

\bibitem{zheng93a}
L. Zheng and A.~H. MacDonald, Phys.\ Rev.\ B {\bf 47},  10619  (1993).

\bibitem{mahan93}
G.~D. Mahan,  in {\em Many-Particle Physics} (Plenum Press, NY, 1993).

\bibitem{wolf85}
E.~L. Wolf,  in {\em Principles of Electron Tunneling Spectroscopy} (Oxford
  University Press, NY, 1985).

\bibitem{eis91b}
J.~P. Eisenstein, T.~J. Gramila, L.~N. Pfeiffer, and K.~W. West, Phys.\ Rev.\ B
  {\bf 44},  6511  (1991).

\bibitem{giuliani82}
G.~F. Giuliani and J.~J. Quinn, Phys.\ Rev.\ B {\bf 26},  4421  (1982).

\bibitem{gold88}
A. Gold, Phys.\ Rev.\ B {\bf 38},  10798  (1988).

\bibitem{coleridge91}
P.~T. Coleridge, Phys.\ Rev.\ B {\bf 44},  3793  (1991).

\bibitem{das93}
B. Das, S. Subramaniam, M.~R. Melloch, and D.~C. Miller, Phys.\ Rev.\ B {\bf
  47},  9650  (1993).

\bibitem{vanhall89}
P.~J. van Hall, Superlatt.\ Microstruct. {\bf 6},  213  (1989).

\bibitem{fuk83}
H. Fukuyama and E. Abrahams, Phys.\ Rev.\ B {\bf 27},  5976  (1983).

\bibitem{note2}
Equation~\ref{e:fuk} contains only the leading term of Eq.~\ref{e:gnq}, and can
  be considered\cite{murp95} to have a prefactor $a_2=\pi^2$.

\bibitem{jung96a}
T. Jungwirth and A.~H. MacDonald, Phys.\ Rev.\ B {\bf 53},  7403  (1996).

\bibitem{zheng96}
L. Zheng and S. {Das Sarma}, Phys.\ Rev.\ B {\bf 53},  9964  (1996).

\bibitem{zhengpriv}
L. Zheng, private communication, (1996).

\bibitem{berk95}
Y. Berk {\it et~al.}, Phys.\ Rev.\ B {\bf 51},  2604  (1995).

\bibitem{slut96}
M. Slutzky {\it et~al.}, Phys.\ Rev.\ B {\bf 53},  4065  (1996).

\bibitem{note1}
Although portions of the two 2DEGs close to the Ohmic contacts have carrier
  densities different to $n_1$ and $n_2$, the resistance of these leads is
  always much smaller than the tunneling resistance, and the results presented
  here depend only on the matched carrier densities in the tunneling region.

\bibitem{eis92a}
J.~P. Eisenstein, L.~N. Pfeiffer, and K.~W. West, Phys.\ Rev.\ Lett. {\bf 69},
  3804  (1992).

\bibitem{smith92}
A.~P. Smith, A.~H. MacDonald, and G. Gumbs, Phys.\ Rev.\ B {\bf 45},  8829
  (1992).

\bibitem{aleiner95a}
I.~L. Aleiner, H.~U. Baranger, and L.~I. Glazman, Phys.\ Rev.\ Lett. {\bf 74},
  3435  (1995).

\bibitem{levi95b}
L.~S. Levitov and A.~V. Shytov, preprint cond-mat/9507058 (unpublished).

\bibitem{aleiner95b}
I.~L. Aleiner and L.~I. Glazman, Phys.\ Rev.\ B {\bf 52},  11296  (1995).

\bibitem{fogler96}
M.~M. Fogler, A.~A. Koulakov, and B.~I. Shklovskii, Phys.\ Rev.\ B {\bf 54},  
1153  (1996).

\bibitem{turn96a}
N. Turner {\it et~al.}, Sol. St. Electron. {\bf 40},  413  (1996).

\end{references}
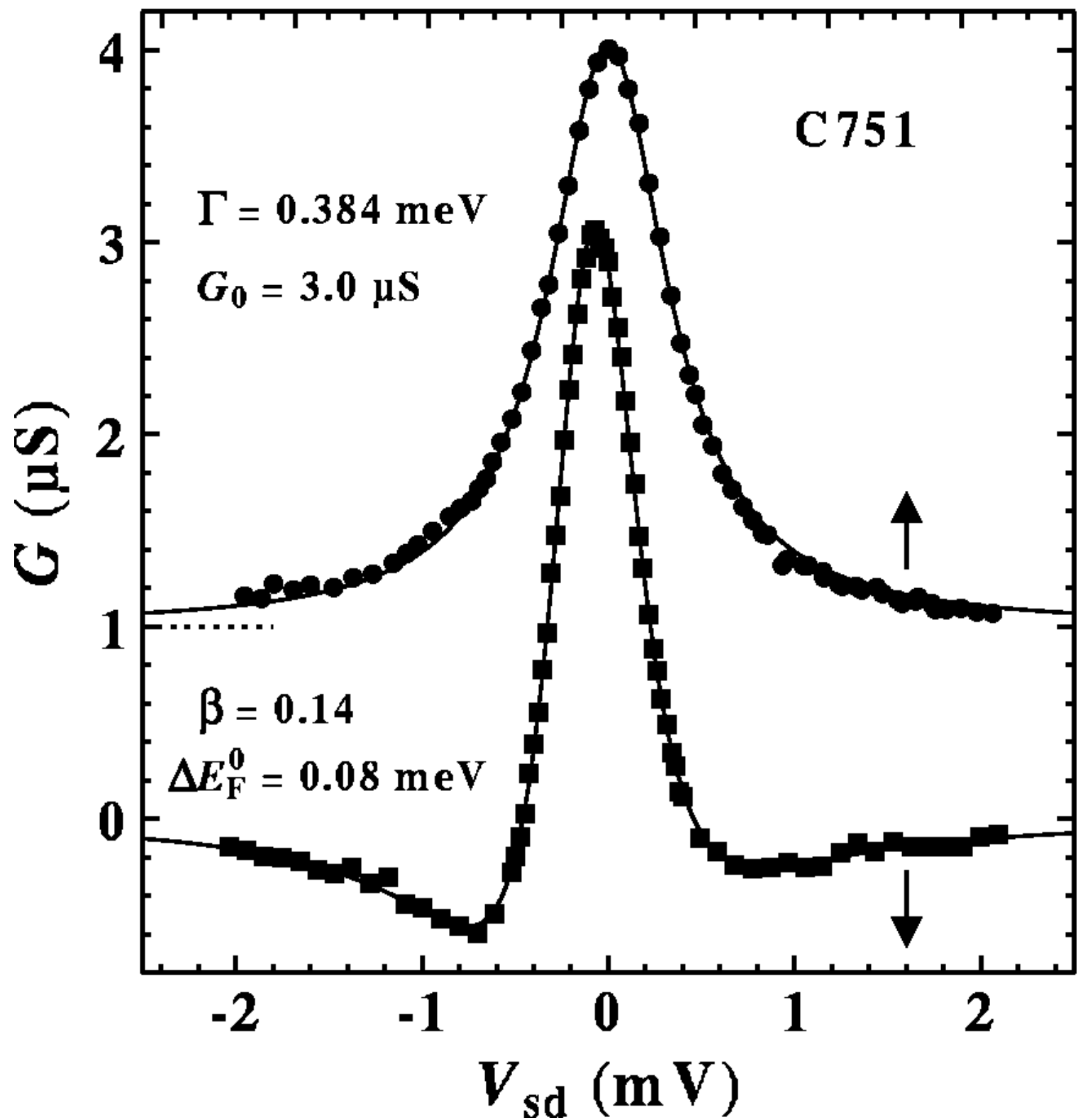
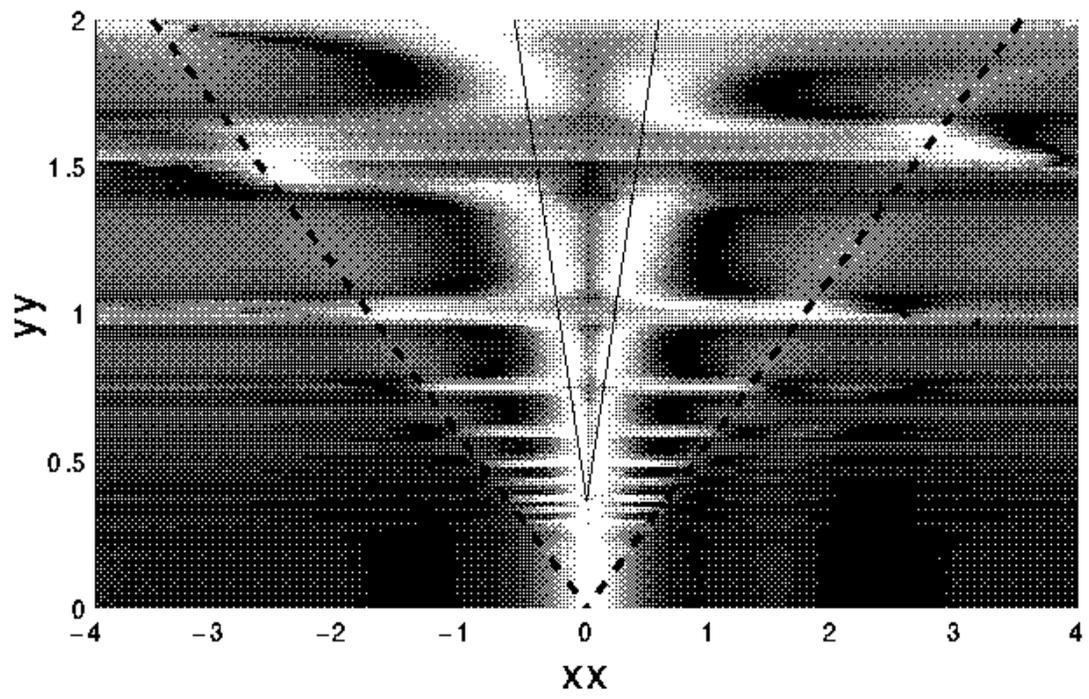
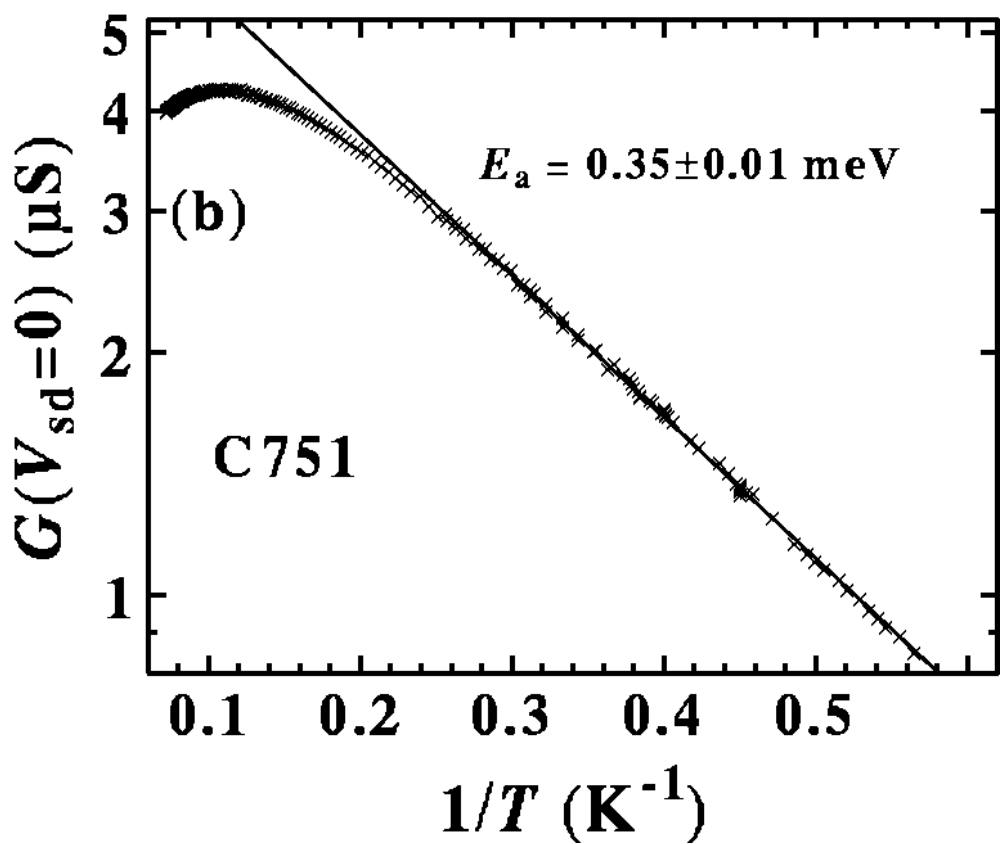
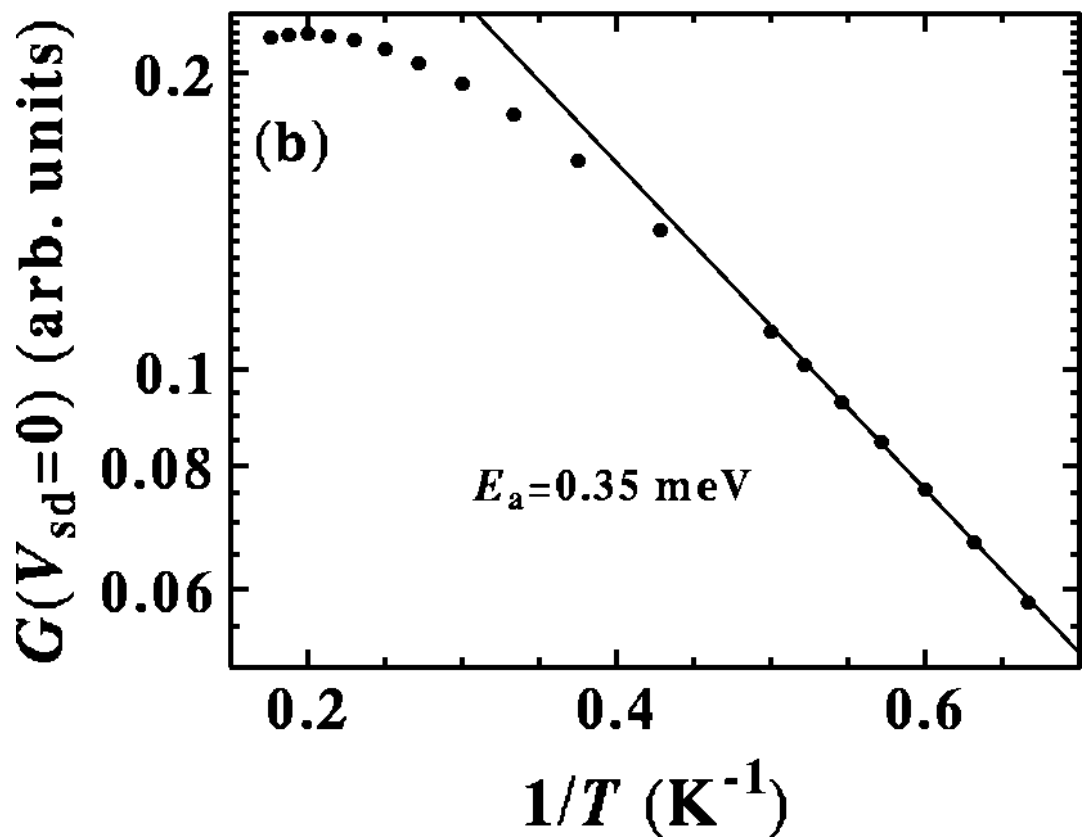
\end{document}